\documentclass[journal]{IEEEtran} 

\usepackage{tabularx} 
 
\usepackage{array} 
\usepackage{multirow} 
\usepackage{colortbl} 
\usepackage{xcolor} 
\usepackage{hyperref} 
\usepackage{makecell} 
\usepackage{booktabs} 
\usepackage{makecell} 
\usepackage{booktabs} 
\usepackage{graphicx} 
\usepackage{tikz} 
\usepackage{tablefootnote} 
\usepackage{xspace}
\usepackage{atbegshi} 
\usepackage{ifthen}  

\usetikzlibrary{ shapes.geometric, arrows.meta, positioning, calc, backgrounds, fit, shapes.symbols, shadows.blur } 
\usepackage[margin=1.5cm]{geometry} 
 
\tikzstyle{startstop} = [circle, draw, fill=black, inner sep=1pt] 
\tikzstyle{process} = [rectangle, draw, minimum height=0.7cm, text centered, rounded corners, font=\small, inner sep=4pt] 
\tikzstyle{decision} = [diamond, draw, aspect=2, text centered, font=\small, inner sep=1pt] \tikzstyle{arrow} = [->,>=stealth, thin] 

\newcommand{\github}[1]{
   \href{#1}{\includegraphics[width=10pt,valign=c]{figs/logos/github-logo.png}}
}

\newcommand{\huggingface}[1]{
   \href{#1}{\includegraphics[width=12pt,valign=c]{figs/logos/hf-logo.png}}
}

\newcommand{\website}[1]{
   \href{#1}{\includegraphics[width=10pt,valign=c]{figs/logos/link_url.png}}
}

\definecolor{navy}{RGB}{0,0,0}

\newcommand{\roham}[1]{\color{black} #1}
\newcommand{\mali}[1]{{\color{black} #1}}
\newcommand{\begum}[1]{{\color{navy} #1}}
\newcommand{\new}[1]{{\color{black}  #1}}

\newcommand{\rohamrev}[1]{{\color{navy}  #1}}

\newcommand{\minorrev}[1]{{\color{navy}  #1}}

\newcommand{\benchcount}{\rohamrev{273}\xspace}
\newcommand{\papercount}{\rohamrev{247}\xspace}
\newcommand{\usercount}{22\xspace}
\newcommand{\modelcount}{ten\xspace}
\newcommand{\searchtool}{BenchScout\xspace}
\newcommand{\benchframework}{BenchFrame\xspace}
\newcommand{\humanevaladvanced}{HumanEvalNext\xspace}
\newcommand{\humanevaladvancedagentic}{HumanEvalNext-Agentic\xspace}

\AtBeginDocument{%
} 

\begin{document} 
\title{ Benchmarking AI Models in Software Engineering:\\ A Review, Search Tool, and Unified Approach for Elevating Benchmark Quality } \author{Roham~Koohestani, Philippe~de~Bekker, Begüm~Koç, and~Maliheh~Izadi
\thanks{R. Koohestani, P. de Bekker, B. Koç, and M. Izadi are with the EEMCS faculty, Delft University of Technology, The Netherlands.}
\thanks{Corresponding author: R. Koohestani (e-mail: rkoohestani@tudelft.nl).} 
\thanks{ORCID: R. Koohestani — 0009-0000-1649-9596; B. Koç — 0009-0000-6686-6008; M. Izadi — 0000-0001-5093-5523.} } %

\maketitle

\begin{abstract}
Benchmarks are essential for unified evaluation and reproducibility. The rapid rise of Artificial Intelligence for Software Engineering (AI4SE) has produced numerous benchmarks for tasks such as code generation and bug repair. However, this proliferation has led to major challenges: (1) fragmented knowledge across tasks, (2) difficulty in selecting contextually relevant benchmarks, (3) lack of standardization in benchmark creation, and (4) flaws that limit utility. Addressing these requires a dual approach: systematically mapping existing benchmarks for informed selection and defining unified guidelines for robust, adaptable benchmark development.
We conduct a review of 247 studies, identifying 273 AI4SE benchmarks since 2014. We categorize them, analyze limitations, and expose gaps in current practices. Building on these insights, we introduce BenchScout, an extensible semantic search tool for locating suitable benchmarks. BenchScout employs automated clustering with contextual embeddings of benchmark-related studies, followed by dimensionality reduction. In a user study with 22 participants, BenchScout achieved usability, effectiveness, and intuitiveness scores of 4.5, 4.0, and 4.1 out of 5.
To improve benchmarking standards, we propose BenchFrame, a unified approach to improve benchmark quality. Applying BenchFrame to HumanEval yielded HumanEvalNext, which features corrected errors, improved language conversion, higher test coverage, and greater difficulty. Evaluating 10 state-of-the-art code models on HumanEval, HumanEvalPlus, and HumanEvalNext revealed average pass-at-1 drops of 31.22\% and 19.94\%, respectively, underscoring the need for continuous benchmark refinement. We further examine BenchFrame's scalability through an agentic pipeline and confirm its generalizability on the MBPP dataset.
Lastly, we publicly release the material of our review, user study, and the enhanced benchmark.~\footnote{\url{https://github.com/AISE-TUDelft/AI4SE-benchmarks}}
\end{abstract}

\section{Introduction}\label{sec:introduction}

\bstctlcite{IEEEexample:BSTcontrol}

Benchmarks are essential for assessing artificial intelligence-driven software engineering (AI4SE) techniques. They provide standardized performance metrics, facilitate reproducibility, and guide innovation. However, the exponential growth in benchmark development has introduced significant challenges: researchers must navigate an increasingly fragmented landscape to identify benchmarks that align well with their specific objectives. This complexity often incentivizes reliance on popular or widely-adopted benchmarks without scrutinizing their applicability, inherent limitations, or potential flaws. Such practices risk propagating biases, overestimating technical progress, and misdirecting research priorities.

\begin{figure}[tb]
    \centering
    \includegraphics[width=0.49\textwidth]{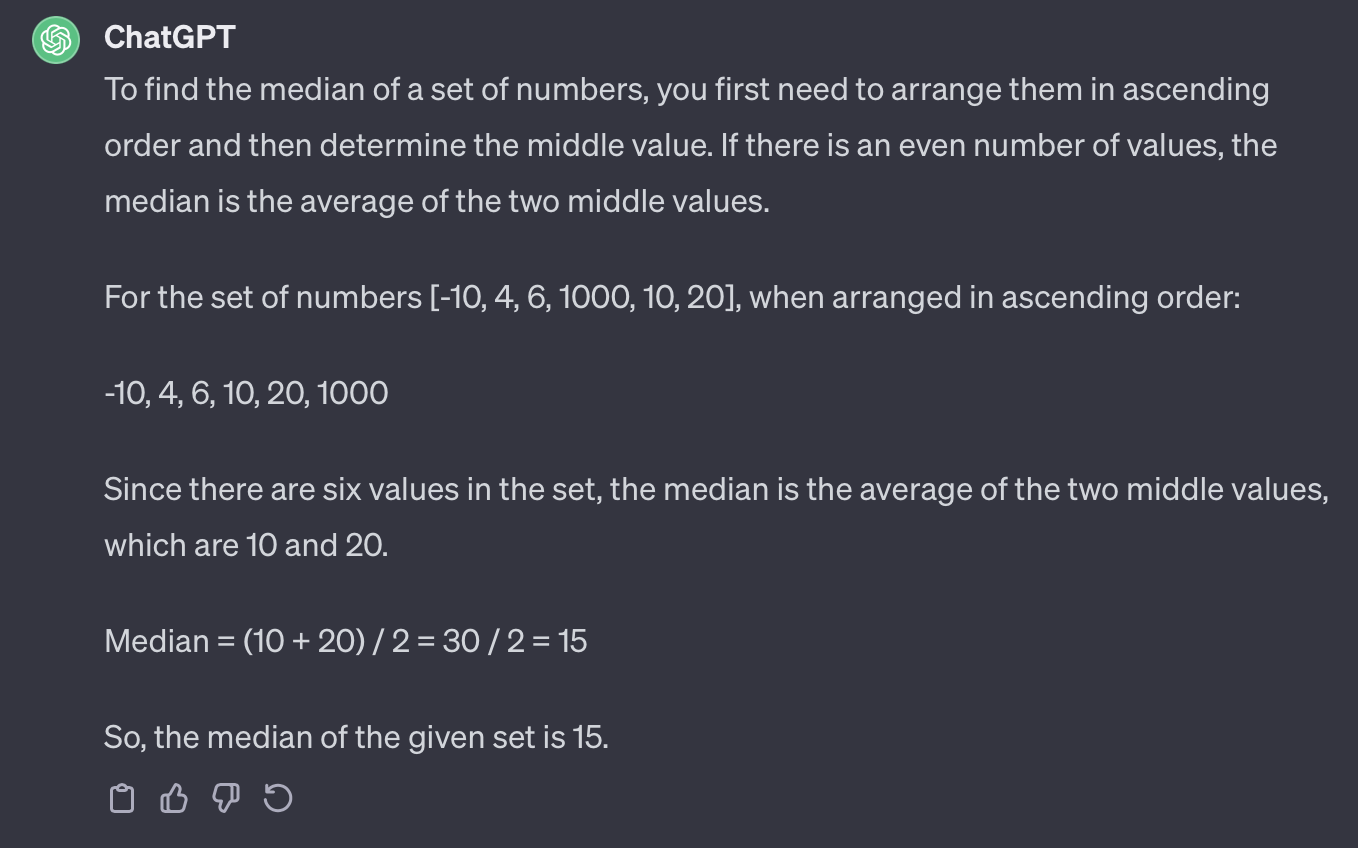}
    \caption{ChatGPT repeats the same HumanEval error (captured on Dec 2023)}
    \label{fig:chatgpt}
\end{figure}
\mali{A notable example of benchmark limitations in code generation evaluation is HumanEval~\cite{HumanEval-2021}, 
a widely-used dataset for assessing large language models (LLMs) such as Codex~\cite{HumanEval-2021}, Gemini~\cite{Gemini-Natural2Code-2023}, and GPT-4~\cite{GPT-4-2023}. 
HumanEval was downloaded 82 thousand times in July 2025 on a single platform (Hugging Face)~\footnote{\url{https://huggingface.co/datasets/openai/openai_humaneval}}, 
reflecting the strong and sustained interest in this benchmark.
Despite its broad adoption, HumanEval contains numerous flaws and inconsistencies~\cite{humaneval-haskell-2024}. 
For instance, Task 47, which requires computing the median of a numerical list, 
incorrectly states the median of the list \texttt{[-10,4,6,1000,10,20]} is $15$. 
When queried on this task, ChatGPT-3.5 Turbo reproduced the incorrect result (see \autoref{fig:chatgpt}).
This suggests potential data contamination and benchmark overfitting which can artificially inflate performance scores.
}

\mali{As HumanEval remains a widely-used benchmark in both AI and software engineering communities, 
several efforts have sought to expand its language support~\cite{MultiEval-HumanEval-MBXP-MathQA-X-2022, MultiPL-E-2022, HumanEvalPack-2023, HumanEval-X-2023} 
or improve test coverage~\cite{CodeScore-HE-APPS-MBPP-Eval-2023, HumanEvalPlus-Mini-MBPP+-2023}.
However, these extensions often build upon the original dataset without addressing its fundamental deficiencies, 
allowing inherent issues, such as flawed canonical solutions, vague problem definitions, incorrect tests, and insufficient coverage to persist. 
Moreover, LLM-augmented improvements, such as automatic translation, often lack rigorous quality control as well.
Lastly, as models have advanced, HumanEval and similar popular benchmarks have become increasingly saturated, 
with close to 100\% scores for recent models.~\footnote{\url{https://evalplus.github.io/leaderboard.html}}
This necessitates continuous elevation of the complexity of programs 
to better reflect the capabilities of contemporary models.
}

\rohamrev{Our central objective is to \textbf{address the systemic challenges within AI4SE benchmarking}. we aim to address four key challenges in benchmarking: (1) the fragmentation of benchmark knowledge across tasks, (2) the difficulty of selecting contextually relevant benchmarks, (3) the lack of standardized approaches for benchmark creation and refinement, and (4) existing inherent flaws that limit benchmark utility. To address this, we use a three-stage approach where each stage builds upon the findings of the previous one.}

More specifically, we first conducted a systematic review of benchmarking efforts in AI4SE from 2014 onward and identified \textbf{\benchcount benchmarks} from \papercount studies. 
\rohamrev{We then extracted a given benchmark's key metadata, 
including (1) objectives, (2) category, (3) programming language, (4) natural language, (5) relevant tasks, (6) extent to which tests are present, (7) scale of the dataset, (8) dataset source, (9) language specificity, (10) maintenance adequacy, (11) whether the dataset is reviewed or not, (12) whether it is frequently used or not, (13) the licensing, and (14) how it was created,
to structure the AI4SE benchmark landscape. 
The findings from our review confirm a fragmented and difficult-to-navigate ecosystem, which motivated us to use these data to develop tooling to easily navigate the landscape of AI4SE benchmarking. \textbf{\searchtool}~\footnote{Accessible through \url{https://evalpro.online/}} is an \textit{extensible semantic} search tool that enables users
to efficiently identify relevant benchmarks for specific software engineering tasks.
}
To build \searchtool, we applied clustering techniques to contextual embeddings derived from related studies and benchmark documentation, 
along with our manually-extracted metadata. 
Additionally, we employed dimensionality reduction techniques to visualize the AI4SE benchmark landscape.
We conducted a user study with \usercount participants from both industry and academia 
to gauge the \textit{usability}, \textit{effectiveness}, and \textit{intuitiveness} of \searchtool. 
It achieved average scores of 4.5, 4.0, and 4.1 out of 5, respectively.

\rohamrev{
While BenchScout aids navigation, our analysis confirms that many foundational benchmarks are flawed, which means a systematic solution is required to deal with them. 
}
Based on the identified gaps and limitations in current benchmarks, we introduce \textbf{\benchframework}, 
a peer-review-oriented methodology to improve the quality of the benchmarks of both existing and new benchmarks. 
To demonstrate its efficacy, we conduct a case study on HumanEval and present \textit{\humanevaladvanced} as an enhanced version.

\mali{When evaluating performance using HumanEval (original) and \humanevaladvanced (our improved version based on the \benchframework), we observe a substantial decline in pass@1 accuracy. Across ten state-of-the-art open-source code models, the average pass@1 score decreases by 31.2\%, with a median drop of 26.0\%. 
Performance remains significantly lower even on HumanEvalPlus~\cite{HumanEvalPlus-Mini-MBPP+-2023}, an enhanced version of HumanEval, with an average decline of $19.94\%$ in pass@1 scores. 
These results highlight the importance of continuously refining benchmarks to better guide future research and provide more realistic assessments of model performance. 
}

\mali{In summary, our contributions are as follows.
\begin{itemize}
    \item We conducted a comprehensive review of \papercount studies, identified \benchcount AI4SE benchmarks, and analyzed their limitations and gaps (\autoref{sec:review}),
    \item We developed and released \textbf{\searchtool}, an extensible semantic search tool to facilitate locating appropriate AI4SE benchmarks. 
    Our user study with \usercount participants demonstrated its effectiveness (\autoref{sec:searchtool}),
    \item We propose a unified approach, \textbf{\benchframework}, to improve the quality and reliability of benchmarks. A case study on HumanEval resulted in a refined, peer-reviewed benchmark, \humanevaladvanced (\autoref{sec:humanevalpro}). We assessed ten recent LLMs on this benchmark and verified significant drops in performance compared to the original HumanEval\cite{HumanEval-2021} and the enhanced version, HumanEvalPlus~\cite{HumanEvalPlus-Mini-MBPP+-2023},
    \item We publicly share our data and details of the user study in our replication package.~\footnote{\url{https://github.com/AISE-TUDelft/AI4SE-benchmarks}}
\end{itemize}}

\rohamrev{
\section{Research Questions}\label{sec:rqs}
To systematically address the challenges outlined in the introduction, we structure our study around our central research objective: \textbf{to diagnose the fragmentation and quality issues within the AI4SE benchmark ecosystem, provide practical tools to navigate this complexity, and establish a rigorous methodology for future quality improvement}. We deconstruct this objective into three guiding Research Questions (RQs), which are designed to be addressed sequentially.

\begin{itemize}
    \item \textbf{RQ1: Diagnosing the problem}: To what extent is the AI4SE benchmarking landscape fragmented, and what are its primary, systemic limitations?

    \item \textbf{RQ2: Providing a Navigational Aid}: How can a semantic search and visualization tool improve the discoverability and selection of relevant AI4SE benchmarks?

    \item \textbf{RQ3: Proposing a Systemic Solution}: What constitutes a rigorous and scalable methodology for improving benchmark quality, and what is its measurable impact on model evaluation?
\end{itemize}

These research questions directly map to the three-stage approach of our study. \textbf{RQ1} fulfills the diagnostic part of our objective by mapping the landscape and identifying its core problems. The findings from RQ1 motivate \textbf{RQ2}, which provides a practical tool to help the community navigate the benchmark fragmentation we uncovered. Finally, \textbf{RQ3} addresses the root cause of the quality issues found in RQ1 by proposing and validating a systematic approach for improvement, thereby fulfilling the final part of our research objective.
}

\section{Existing Benchmarks, a Review}
\label{sec:review}
\subsection{Search Criteria and Quality Assessment}

We employed a systematic method to search for, identify, and classify AI4SE benchmarks. This procedure involved three primary stages: conducting structured searches, verifying credibility, and developing a taxonomy. We searched on two platforms, namely Google Scholar and Semantic Scholar, using the following keywords: ``Benchmarks'', ``Software Engineering'', ``Large Language Models'', ``Evaluation'', and ``AI4SE''. We additionally searched for benchmarks present in the PapersWithCode datasets collection due to the popularity and wide usage of the platform.~\footnote{This platform has since been discontinued and is no longer accessible. The previously used URL was \url{https://paperswithcode.com/datasets}}

We selected these keywords based on a preliminary assessment of highly-cited benchmark research. Three authors worked together to revise the search criteria to ensure they were both relevant and complete. Our selection criteria targeted primary studies published in English from 2014 to 2025.
After the initial paper collection process, duplicates were removed. Two authors reviewed the relevance of the identified pieces of literature, during which the originality (that is, the status as primary study), reproducibility, and accessibility of each study were accessed. 

More specifically, for originality, we assessed whether a study was a primary study that introduced a new benchmark or a significant extension of an existing one. This filter was set in place to distinguish between benchmark-proposing papers from studies that merely applied them with a certain level preprocessing. In terms of Reproducibility and Accessibility, we evaluated whether the benchmark and its associated resources were made publicly available. We looked for the presence of a public repository (e.g., GitHub or HuggingFace) that would allow the community to access and use the benchmark.
After this phase, we performed forward and backward snowballing. A total of \papercount papers were found through this process.

After the papers were collected, the authors worked together to develop and consistently improve a taxonomy to efficiently categorize the benchmarks and extract metadata. Through continuous discussions, we identified initial essential categories while systematically gathering additional details for each study, including DOI and publication date. An iterative strategy was used to design the taxonomy and categorize SE tasks, starting with overarching categories such as reasoning, synthesis, and debugging. When the categories became too broad or lacked cohesion, the authors refined them into more detailed subcategories, establishing a multi-tiered hierarchy. \minorrev{
Furthermore, it is important to highlight that during the categorization process, the two authors involved discovered 10 instances of disagreement over category classifications, resulting in an agreement rate of 96.4\%. This outcome can be attributed to the relatively objective nature of benchmark categorization. Many of the papers either clearly specify their intended task or imply it within the text, with the latter being the source of discrepancies in the few instances where the raters disagreed.
}

In our evaluation, we not only considered benchmark datasets from academic sources, but also included those suggested by industry and subsequently adopted by scholars (for example, IBM CODAIT and Aider~\cite{CODAIT-2021}). Our replication package also serves as a dynamic repository, allowing researchers to contribute additional benchmarks and related papers by submitting a pull request that includes the paper's DOI.

\rohamrev{
After finalizing the taxonomy, and the list of benchmarks, we extracted (1) objectives, (2) categories, (3) programming language, (4) natural languages, (5) relevant tasks, (6) extent to which tests are present, (7) scale of the dataset, (8) dataset source, (9) language specificity, (10) maintenance adequacy, (11) whether the dataset is reviewed or not, (12) whether it is frequently used or not, (13) the licensing, and (14) how it was created, for each of the benchmarks.
}

\subsection{Results of the Review}
\begin{figure}
    \centering
    \includegraphics[width=0.49\textwidth]{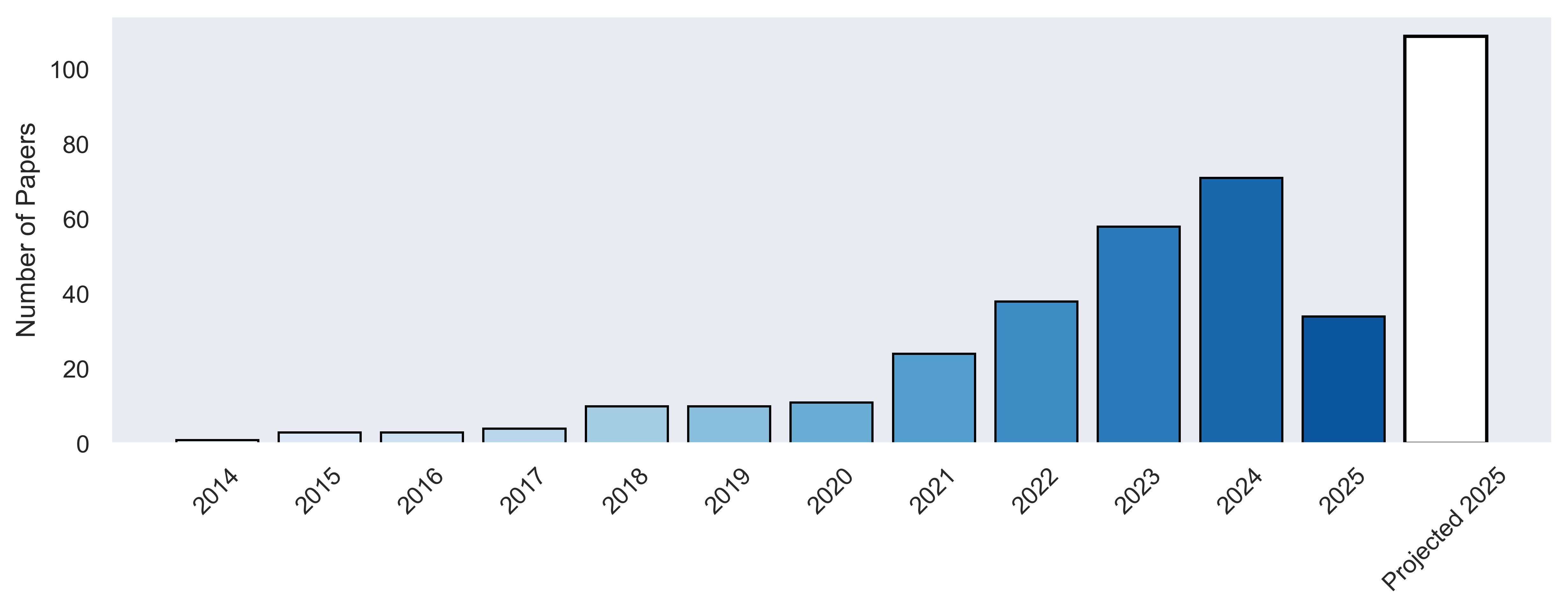}
    \caption{\rohamrev{Number of published benchmark papers (Jan 2014-Aug 2025).}}
    \label{fig:paper_count}
\end{figure}

Our review revealed a significant increase in the number of benchmarks published over time (see \autoref{fig:paper_count}). We identified \benchcount benchmarks in total, with 71 published in 2024 alone. Using an exponential projection, we estimate that this number will reach 109 for the year 2025. This highlights the growing impact of AI4SE benchmarking and the need for a comprehensive overview of the existing literature. 

Additionally, upon analyzing the distribution of the tasks included in the review, we find that a great portion of benchmarks are related to \textit{Code Generation}, more specifically, 34.4\% of all the benchmarks included. Along with Code Understanding and Repair \& Maintenance (with respectively 17.7\% and 15.4\%), these make up more than 67\% of all benchmarks included. This, in itself, underscores a lack of attention to tasks such as vulnerability detection and code retrieval/search (cf. \autoref{fig:category_distribution}). 

\begin{figure}
    \centering
    \includegraphics[width=0.49\textwidth]{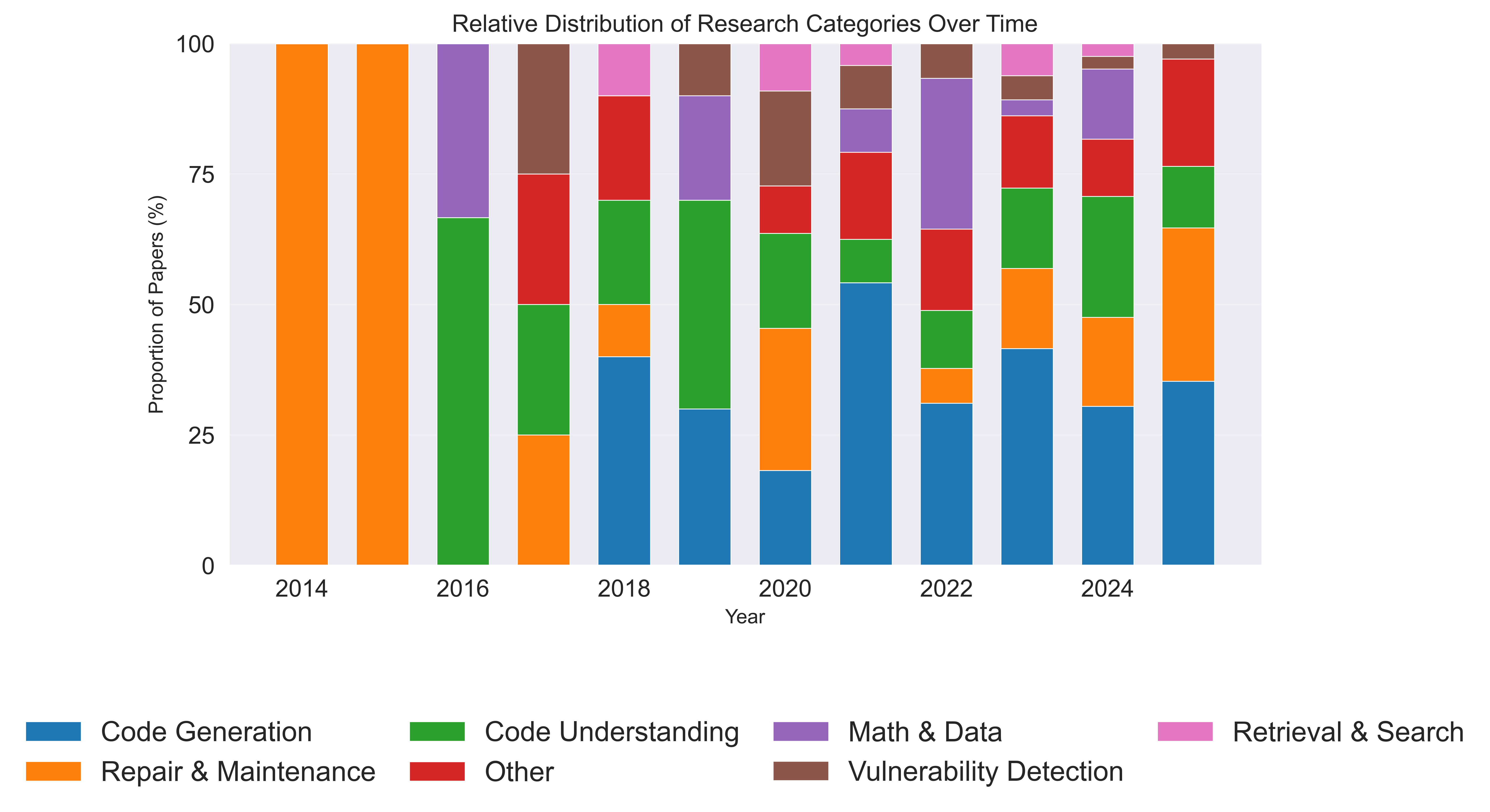}
    \caption{\rohamrev{Task distribution progression for the benchmarks included in the study.}}
    \label{fig:category_distribution_time}
\end{figure}

The data reveal a notable change over time in the distribution of research categories throughout the observed period (cf. \autoref{fig:category_distribution_time}). A key insight is the shift from code repair and maintenance to automated synthesis, marked by two opposing trends. Initially dominant, the Repair \& Maintenance category gradually experiences a notable decline in its share. In contrast, starting around 2020, the Code Generation category gains prominence, which coincides with the advancement of large-scale language models. In the recent years presented, there is an emerging trend of thematic diversification; as the focus on Code Generation recedes from its high, categories like Code Understanding and Retrieval \& Search show a relative resurgence. This suggests a community-wide transition towards tackling additional challenges related to verification, integration, and the understanding of algorithmically generated outputs. In the following, we summarize key insights from our review.

\begin{figure}[tb]
\centering
\begin{tikzpicture}[font=\small, line width=0.3pt, scale=0.8, every node/.append style={transform shape}]
  \def\W{10.0}   
  \def\H{6.0}    

  \newcommand{\treerect}[9]{%
    \fill[#5, rounded corners=2pt] (#1,#2) rectangle ++(#3,#4);
    \draw[white, rounded corners=2pt] (#1,#2) rectangle ++(#3,#4);
    \node[text=black, align=center, text width=#7, inner sep=1pt,
          font=\fontsize{#8}{#8}\selectfont, rotate=#9]
         at ({#1 + #3/2},{#2 + #4/2}) {#6};
  }

  \definecolor{cCodeGen}{RGB}{31,119,180}     
  \definecolor{cCodeUnd}{RGB}{44,160,44}      
  \definecolor{cRepair}{RGB}{255,127,14}      
  \definecolor{cOther}{RGB}{214,39,40}        
  \definecolor{cMath}{RGB}{148,103,189}       
  \definecolor{cVuln}{RGB}{140,86,75}         
  \definecolor{cRetrieval}{RGB}{227, 119, 194}  

  \treerect{0.000}{0.000}{3.440}{6.000}{cCodeGen}{Code Generation\\(34.4\%)}{2.9cm}{8.6}{0}

  \def\xR{3.440}   \def\wR{6.560}

  \treerect{\xR}{0.000}{\wR}{1.620}{cCodeUnd}{Code Understanding\\(17.7\%)}{6.1cm}{7.6}{0}

  \treerect{\xR}{1.620}{2.110}{4.380}{cRepair}{Repair \&\\Maintenance\\(15.4\%)}{3.5cm}{7.6}{0}

  \treerect{5.550}{1.620}{4.450}{1.810}{cOther}{Other\\(13.4\%)}{4.1cm}{7.6}{0}

  \treerect{5.550}{3.430}{2.500}{2.570}{cMath}{Math \& Data\\(10.7\%)}{3.0cm}{7.6}{0}

  \treerect{8.050}{3.430}{1.950}{1.660}{cVuln}{Vulnerability\\Detection\\(5.4\%)}{1.8cm}{7.2}{0}

  \treerect{8.050}{5.090}{1.950}{0.910}{cRetrieval}{Retrieval \&\\Search\\(3.0\%)}{1.75cm}{6.8}{0}

\end{tikzpicture}
\caption{High-level category distribution of included benchmarks (updated, alternating orientation).}
\label{fig:category_distribution}
\end{figure}

\footnotesize
\begin{table*}[tb]
  \caption{Overview of AI4SE benchmarks stemming from HumanEval~\cite{HumanEval-2021} (PL denotes Programming Language, NL denotes Natural Language).}
  \label{tab:humaneval_fam}
  \centering
  \begin{tabular}{lllc}
    \toprule
    \textbf{Category} & \textbf{Name} & \textbf{Language(s)} & \textbf{\# Tests} \\
    \addlinespace[0.5ex]
    \midrule
    Original & HumanEval~\cite{HumanEval-2021} & Python & Avg. 7.7 \\
    \addlinespace[0.5ex]
    \midrule
    \multirow{8}{*}{Improved Language Support} 
    & MultiPL-HumanEval~\cite{MultiPL-E-2022} & 18 PLs & Avg. 7.7 \\
    & HumanEval-Fix~\cite{HumanEvalPack-2023} & 6 PLs & Avg. 7.7 \\
    & HumanEval-Explain~\cite{HumanEvalPack-2023} & 6 PLs & Avg. 7.7 \\
    & HumanEval-Synthesize~\cite{HumanEvalPack-2023} & 6 PLs & Avg. 7.7 \\
    & HumanEval-X~\cite{HumanEval-X-2023} & 5 PLs & Avg. 7.7 \\
    & Multi-HumanEval~\cite{MultiEval-HumanEval-MBXP-MathQA-X-2022} & 12 PLs & Avg. 7.7 \\
    & HumanEvalXL~\cite{peng_humaneval-xl_2024} & 12 PLs, 23 NLs & Avg. 8.33 \\
    & \begum{mHumanEval~\cite{raihan2024mhumaneval}} & \begum{Python, 204 NLs} & \begum{Avg. 7.7} \\
    \addlinespace[0.5ex]
    \midrule
    \multirow{3}{*}{Improved Testing}
    & HumanEval+~\cite{HumanEvalPlus-Mini-MBPP+-2023} & Python & Scaled ×80 \\
    & HumanEval-MINI~\cite{HumanEvalPlus-Mini-MBPP+-2023} & Python & Scaled ×47 \\
    & HE-Eval~\cite{CodeScore-HE-APPS-MBPP-Eval-2023} & Python & Scaled ×14 \\
    \addlinespace[0.5ex]
    \midrule
    Instruction-based & InstructHumanEval~\tablefootnote{\url{https://huggingface.co/datasets/codeparrot/instructhumaneval}} & Python & Avg. 7.7 \\
    \addlinespace[0.5ex]
    \midrule
    \multirow{2}{*}{Extended} & \multirow{2}{*}{EvoEval~\cite{EvoEval-2024}} & \multirow{2}{*}{Python} & Multiple categories, \\ & & &  scaled with \textsc{EvalPlus} \\
    \addlinespace[0.5ex]
    \bottomrule
  \end{tabular}
\end{table*}

\normalsize

\textbf{HumanEval Benchmark Family}: 
Currently, one of the most popular AI4SE benchmarks is HumanEval~\cite{HumanEval-2021}, used to evaluate the performance of many notable code-aware models (e.g., Codex~\cite{HumanEval-2021}, Gemini~\cite{Gemini-Natural2Code-2023}, and GPT-4~\cite{GPT-4-2023}). This benchmark is used mainly for code synthesis, though there exist some variations for code repair and code explanation~\cite{HumanEvalPack-2023}. 
\autoref{tab:humaneval_fam} presents the family of HumanEval benchmarks. After an in-depth analysis of these benchmarks, we identified the following issues: (1) incorrect tests, (2) lack of proper test coverage, (3) incorrect canonical solutions, and (4) imprecise problem definitions. While there are versions that have improved the language support~\cite{MultiEval-HumanEval-MBXP-MathQA-X-2022, MultiPL-E-2022, HumanEvalPack-2023, HumanEval-X-2023} and test coverage~\cite{CodeScore-HE-APPS-MBPP-Eval-2023, HumanEvalPlus-Mini-MBPP+-2023}, there is no version that contains all the improvements combined nor fixed the original issues. The issues for enhancing the original dataset can be generalized as follows:
\begin{itemize}
    \item Variants that cover multiple languages have duplicated the original issues.
    \item Variants that added tests used the original incorrect solutions to generate the output.
    \item Variants based on human corrections or translations are inconsistent.
\end{itemize}

Furthermore, production systems like ChatGPT-3.5 tend to replicate errors from the original HumanEval benchmark. This indicates potential contamination from the benchmark data, \textit{with the systems not only producing incorrect answers but also seemingly optimizing to match the flawed responses from the widely used benchmark}. Given the widespread use and ongoing popularity of the HumanEval benchmark in the research community, it is crucial to address its inherent flaws to prevent the perpetuation of these issues.

\footnotesize
\begin{table}[tb]
  \caption{Overview of AI4SE benchmarks derived from MBPP~\cite{MBPP-MathQA-2021}.}
  \label{tab:mbpp_fam}
  \centering
  \begin{tabular}{p{12mm}llp{20mm}}
    \toprule
    \textbf{Category} & \textbf{Name} & \textbf{Language(s)} & \textbf{\# Problems} \\
    \addlinespace[0.5ex]
    \midrule
    Original & MBPP~\cite{MBPP-MathQA-2021} & Python & 974 \\
    \addlinespace[0.5ex]
    \midrule
    Improved Language & MultiPL-MBPP~\cite{MultiPL-E-2022} & 18 PLs & 354-397 per PL\\
    Support & MBXP~\cite{MultiEval-HumanEval-MBXP-MathQA-X-2022} & 13 PLs & 848-974 per PL \\
    \addlinespace[0.5ex]
    \midrule
    Improved & MBPP+~\cite{HumanEvalPlus-Mini-MBPP+-2023} & Python & 427 \\
    Testing & MBPP-Eval~\cite{CodeScore-HE-APPS-MBPP-Eval-2023} & Python & 974 \\
    \addlinespace[0.5ex]
    \bottomrule
  \end{tabular}
\end{table}
\normalsize

\textbf{MBPP Benchmark Family}:
 Another AI4SE benchmark, highly similar in style and popularity compared to HumanEval, is MBPP~\cite{MBPP-MathQA-2021}: Mostly Basic Python Problems. It contains nearly a thousand crowdsourced problems, where almost half of it is sanitized and separately released. Furthermore, several enhancements have been published for MBPP (\autoref{tab:mbpp_fam}). Upon a more in-depth analysis of MBPP and its family of benchmarks, there are many signs suggesting deficient quality. One notable problem is the lack of proper testing, as MBPP originally only has three (rather trivial) tests per problem, which are all revealed in the prompt as well. With such a test suite in place, evaluation metrics become unstable and insignificant for proper comparison. The strength of the written tests and solutions themselves is not only troublesome in the original data but also the sanitized data features many flaws (even in \textit{corrected} variants~\cite{HumanEvalPlus-Mini-MBPP+-2023}). From negligible observations such as poor syntax (e.g., too many spaces, Python method names starting with a capital -- this is a common convention to only use for classes) to uncaught bugs and edge cases that break the implementation. While there are enhancements that improve the language support and extend the test cases, they are all built upon inadequate foundations, which renders any MBPP benchmark suboptimal for proper assessment. 

\textbf{Other Existing Benchmarks}:
 Besides HumanEval and MBPP, the standardized benchmarks for code synthesis evaluation, there are many more considerable benchmarks for assessing various categories of SE tasks. \mali{We share several additional categorized tables as a guide for finding specific AI4SE benchmarks and highlight the most notable benchmarks for each in detail}. 
 
\footnotesize
\begin{table*}[tb]
  \caption{Overview of competitive programming, code complexity, and code efficiency benchmarks.}
  \label{tab:table_cp_cc_ce}
  \centering
  \begin{tabular}{lllc}
    \toprule
    \textbf{Category} & \textbf{Name} & \textbf{Language(s)} & \textbf{\# Tests} \\
    \addlinespace[0.5ex]
    \midrule
    \multirow{10}{*}{Competitive Programming} 
    & CodeContests~\cite{CodeContests-AlphaCode-2022} & 12 PLs & Avg. 203.7 \\
    & APPS~\cite{APPS-2021} & Python & Avg. 13.2 \\
    & LiveCodeBench~\cite{jain_livecodebench_2024} & Python & Avg. 17.23 \\
    & LeetCode~\cite{tian_is_2023} & Python & Avg. 135 \\
    & CodeElo~\cite{quan_codeelo_2025} & N/A & 408 problems \\
    & \begum{FVAPPS~\cite{dougherty2025proving}} & \begum{Python} & \begum{4715 problems}\\
    & \begum{KodCode~\cite{xu2025kodcode}} & \begum{12 PLs} & \begum{Avg. 7.52}\\
    & \begum{FC2Code~\cite{liu2022code}} & \begum{Python} & \begum{320 flowcharts} \\
    & \begum{CodeContests+~\cite{wang2025codecontests}} & \begum{Python} & \begum{11,690 problems} \\ 
    & \begum{LiveCodeBenchPro~\cite{zheng2025livecodebench}} & \begum{Python} & \begum{584 problems} \\
    \addlinespace[0.5ex]
    \midrule
    \multirow{5}{*}{Code Complexity} 
    & CoRCoD~\cite{CoRCoD-2019} & Java & 932 \\
    & GeeksForGeeks (GFG)~\cite{GFG-2023} & C++, Python & ±1,400 per lang.\&categ \\
    & CODAIT~\tablefootnote{CODAIT-2021 \url{https://ibm.co/4emPBIa}} & Python & 4,000,000 \\
    & CodeComplex~\cite{CodeComplex-2022} & Java, Python & 4,900 per language \\
    & PythonSaga~\cite{PythonSaga-HumanEval-MBPP-Evaluation-Difficulty-2024} & Python & 185 \\
    \addlinespace[0.5ex]
    \midrule
    \multirow{4}{*}{Code Efficiency} & EffiBench~\cite{EffiBench-2024} & Python & Self-defined, avg. 100 \\
    & CODAL~\cite{m_weyssow_codeultrafeedback_2024} & Python & 3 ref. / problem \\
    & PIE~\cite{shypula_learning_2024} & C++ & 82.5(median, train)\\
    & \begum{COFFE~\cite{peng2025coffe}} & \begum{Python} & \begum{756} \\
    
    \addlinespace[0.5ex]
    \bottomrule
  \end{tabular}
\end{table*}
\normalsize

\footnotesize
\begin{table*}[tb]
  \caption{Overview of data science \& domain-specific benchmarks. (JN refers to Jupyter Notebooks.)}
  \label{tab:table_ds}
  \centering
  \begin{tabular}{lllc}
    \toprule
    \textbf{Name} & \textbf{Language(s)} & \textbf{\# Tests} & \textbf{Comment} \\
    \addlinespace[0.5ex]
    \midrule
    DS-1000~\cite{DS-1000-2022} & Python & Avg. 1.6 & 7 DS/ML libraries \\
    \multirow{1}{*}{NumpyEval~\cite{NumpyEval-PandasEval-PyCodeGPT-2022}} & \multirow{1}{*}{Python} & \multirow{1}{*}{Avg. 20 functions} & NumPy (101 problems) \\ 
    & & & (Avg. 1 variable) \\
    \multirow{1}{*}{PandasEval~\cite{NumpyEval-PandasEval-PyCodeGPT-2022}} & \multirow{1}{*}{Python} & \multirow{1}{*}{Avg. 20 functions} & Pandas (101 problems) \\ 
    & & & (Avg. 1 variable) \\
    \multirow{1}{*}{JuICe~\cite{JuICe-2019}} & \multirow{1}{*}{Python, JN} & \multirow{1}{*}{N/A} & Cell completion \\ 
    & & & (1.5M/3.7K train test) \\
    \multirow{1}{*}{DSP~\cite{DSP-2022}} & \multirow{1}{*}{Python, JN} & \multirow{1}{*}{Available} & Cell completion \\ 
    & & & (1,119 problems) \\
    \multirow{1}{*}{ExeDS~\cite{huang_junjie_execution-based_2022}} & \multirow{1}{*}{Python, JN} & \multirow{1}{*}{Execution Based} & Cell generation \\ 
    & & & (ground truth), 534 tasks \\
    \multirow{1}{*}{DSEval~\cite{zhang_benchmarking_2024}} & \multirow{1}{*}{Python} & \multirow{1}{*}{custom appraoch} & Models Evaluated via the DSEval \\
    & & & Approach from the Paper \\
    \multirow{1}{*}{\begum{TorchDataEval}~\cite{zan2022language}} & \multirow{1}{*}{\begum{Python}} & \multirow{1}{*}{\begum{50}} & \begum{Private PyTorch data library} \\
    \multirow{1}{*}{\begum{MonkeyEval}~\cite{zan2022language}} & \multirow{1}{*}{\begum{Python}} & \multirow{1}{*}{\begum{101}} & \begum{Private Pandas library fork} \\
    \multirow{1}{*}{\begum{BeatNumEval}~\cite{zan2022language}} & \multirow{1}{*}{\begum{Python}} & \multirow{1}{*}{\begum{101}} & \begum{Private NumPy variant} \\

    \midrule
    \multirow{2}{*}{Bio-Coder~\cite{xiangru_tang_biocoder_2024}} & Python & 1,026 & Identify and import necessary \\
    & Java & 1,243 & classes for given task \\
    \multirow{1}{*}{Bio-Coder-Rosalind~\cite{xiangru_tang_biocoder_2024}} & \multirow{1}{*}{Python} & 253 golden solution & Generate code for question  \\ 
    \multirow{1}{*}{WebApp1k~\cite{cui2024webapp1kpracticalcodegenerationbenchmark}} & \multirow{2}{*}{React} & \multirow{1}{*}{Available} & evaluates whether a model can \\
    & & & generate React web-app\\
    \addlinespace[0.5ex]
    \bottomrule
  \end{tabular}
\end{table*}
\normalsize
\footnotesize
\begin{table}[tb]
  \caption{Overview of Mathematical Reasoning Benchmarks.}
  \label{ref:table_math}
  \centering
  \begin{tabular}{llp{25mm}}
    \toprule
    \textbf{Name} & \textbf{Language(s)} & \textbf{\# Problems} \\
    \addlinespace[0.5ex]
    \midrule
    MATH~\cite{dan_hendrycks_measuring_2021} & English & 12,500 \\
    MATH500~\cite{lightman_lets_2023} & English & 500 \\
    MathQA~\cite{MathQA-2019} & English & 37,297 \\
    MathQA-Python~\cite{MBPP-MathQA-2021} & Python & 23,914 \\
    MathQA-X~\cite{MultiEval-HumanEval-MBXP-MathQA-X-2022} & Python, Java, JS & 1,883 per language \\
    \textsc{LiLa}\cite{mishra_swaroop_lila_2022} & Python & 133,815 questions \\
    & & 358,769 programs \\
    MultiArith~\cite{MultiArith-2015} & English & 600 \\
    GSM8K~\cite{PAL-GSM-Math-2022} & English & 1,320 \\
    GSM-HARD~\cite{PAL-GSM-Math-2022} & English & 1,320 \\
    TheoremQA~\cite{TheoremQA-2023} & English & 800 \\
    PECC~\cite{patrick_haller_pecc_2024} & Python & 1,006 \\
    BRIGHT~\cite{hongjin_su_bright_2024} & English & 395 \\
    AMC12\footnote{\href{https://huggingface.co/datasets/AI-MO/aimo-validation-amc}{https://huggingface.co/datasets/AI-MO/aimo-validation-amc}} & English & 82 \\

    \addlinespace[0.5ex]
    \bottomrule
  \end{tabular}
\end{table}
\normalsize

\footnotesize
\begin{table}[tb]
  \caption{Overview of Natural Language Benchmarks. These are specifically the Text2Code Bencharmks.}
  \label{ref:table_nl}
  \centering
  \begin{tabular}{llc}
    \toprule
    \textbf{Name} & \textbf{Language(s)} & \textbf{\# Problems} \\
    \addlinespace[0.5ex]
    \midrule
    CoNaLa~\cite{CoNaLa-2018} & English $\rightarrow$ Python & 2,879 \\
    MCoNaLa~\cite{MCoNaLa-2022} & \{Spanish, Japanese, & 896 \\
    & Russian\} $\rightarrow$ Python & \\
    CoNaLa-SO~\cite{orlanski_reading_2021} & English $\rightarrow$ Python & 10,000\\
    APPS~\cite{APPS-2021} & English $\rightarrow$ Python & 10,000 \\
    APPS-Eval~\cite{CodeScore-HE-APPS-MBPP-Eval-2023} & English $\rightarrow$ Python & 10,000 \\
    AixBench~\cite{AixBench-2022} & English, Chinese & 175 \\
    & $\rightarrow$ Java & \\
    Natural2Code~\cite{Gemini-Natural2Code-2023} & English $\rightarrow$ Python & Unknown \\
    CoSQA~\cite{CoSQA-2021} & English $\rightarrow$ Python & 20,604 \\
    WebQueryTest~\cite{CodeXGLUE-2021} & English $\rightarrow$ Python & 1,046 \\
    AdvTest~\cite{CodeXGLUE-2021} & English $\rightarrow$ Python & 280,634 \\
    CONCODE~\cite{Concode-2018} & English $\rightarrow$ Java & 104,000 \\
    MTPB~\cite{MTPB-CodeGen-2022} & English $\rightarrow$ Python & 115 \\
    CAASD ~\cite{simiao_zhang_experimenting_2024} & English $\rightarrow$ Python & 72\\
    Shellcode\_IA32 ~\cite{liguori_can_2022} & English $\rightarrow$ IA32/Shell & 3200 \\
    Odex ~\cite{liguori_can_2022} & \{Spanish, Japanese,  & 945 \{90, 164,  \\
    & Russian, English\} & 252, 439\} \\
    & $\quad\quad\quad\rightarrow$ Python & 1707 test total\\
    PSB2 ~\cite{thomas_helmuth_psb2_2021} & English $\rightarrow$  & 25 \\
    & \{Clojure, Python\} & question-answer pairs\\
    TACO ~\cite{rongao_li_taco_2023} & English $\rightarrow$ Python & 1,539,152 on 26,433 \\
    & &  distinct tasks\\
    Turbulence ~\cite{shahin_honarvar_turbulence_2023} & English $\rightarrow$ Python & 60 (with 420 \\
    &  & total test cases)\\
    Aider~\tablefootnote{\href{https://github.com/Aider-AI/aider/blob/main/benchmark/README.md}{https://github.com/Aider-AI/aider/blob/main/benchmark/README.md}} & English  & \\
    & $\rightarrow$ \{C++, GO, Java,  & 225 problmes \\
    & JS, Python, Rust\} & \\
    NL2ML-lib~\cite{shin_good_2024} & English $\rightarrow$ Python & 11,000 \\ 
    RMCBench~\cite{chen_rmcbench_2024} & English & 473 malicious  \\
    & $\rightarrow$ 9 Languages & prompts \\
    Evil~\cite{liguori_evil_2021} & English & 19255 \\
    & $\rightarrow$ \{Python, IA\_32\} & \\
    \begum{Exec-CSN~\cite{xie2024codebenchgen}} & \begum{English $\rightarrow$ Python} & \begum{1,931} \\
    \begum{CodeIF~\cite{yan2025codeif}} & \begum{English $\rightarrow$ } & \begum{1,200 tasks} \\
    & \{Java, Python, & \\
    &  Go, C++\} & \\
    \begum{CodeIF-Bench~\cite{wang2025codeif}} & \begum{English $\rightarrow$ Python} & \begum{122 tasks} \\
    & & \begum{(42 repositories)} \\
    \begum{ARCADE~\cite{yin2023natural}} & \begum{English $\rightarrow$ Python} & \begum{1082} \\
    \begum{StackEval~\cite{shah2024stackeval}} & \begum{English $\rightarrow$ Python} & \begum{925} \\
    \begum{SwiftEval~\cite{petrukha2025swifteval}} & \begum{English $\rightarrow$ Swift} & \begum{28} \\
    \begum{CoSQA+~\cite{gong2025cosqapioneeringmultichoicecode}} & \begum{English $\rightarrow$ Python} & \begum{412,080 pairs} \\

    \addlinespace[0.5ex]
    \bottomrule
  \end{tabular}
\end{table}

\begin{table}[tb]
  \caption{Overview of Natural Language Benchmarks (Continued). The First two benchmarks are Text2Text (related to code) benchmarks, while the rest are Code2Text benchmarks.}
  \label{ref:table_nl2}
  \centering
  \begin{tabular}{llc}
    \toprule
    \textbf{Name} & \textbf{Language(s)} & \textbf{\# Problems} \\
    \addlinespace[0.5ex]
    \midrule 
    InfiCoder-Eval~\cite{InfiCoder-Eval-2023} & English $\rightarrow$ English & 270 \\
    BRIGHT~\cite{hongjin_su_bright_2024} & English $\rightarrow$ English & 1,398 \\
    \midrule
    DeepCom~\cite{DeepCom-2018} & Java $\rightarrow$ English & 588K \\
    Hybrid-DeepCom~\cite{hu_deep_2020} & Java $\rightarrow$ English & 466k \\
    BinSum~\cite{BinSum-2023} & Binary functions & 557K \\
     & $\rightarrow$ English &  \\
    Code Attention ~\cite{miltiadis_allamanis_convolutional_2016} & Java $\rightarrow$ English & 11 projects \\ 
    Funcom ~\cite{alexander_leclair_neural_2019} & Java $\rightarrow$ English & 2.1M problems \\ 
    CodeSum~\cite{hu_summarizing_nodate} & Java $\rightarrow$ English & 410,630 \\
    CoDesc~\cite{hasan_codesc_2021} & Java $\rightarrow$ English & 4.21M datapoints \\
    Parallel~\cite{barone_parallel_nodate} & Python $\rightarrow$ English & 150k function/doc pais \\
    CoDocBench~\cite{pai2025codocbenchdatasetcodedocumentationalignment} &  Python $\rightarrow$ English & 4573 code/doc pairs \\
    \begum{PoorCodeSum~\cite{hu2024effectively}} & \begum{\{Java, Python, Go\} } & \begum{\{10,955 , 14,918} \\
    & \begum{$\rightarrow$ English} &  , 8,122\} \\
    \begum{P-CodeSum~\cite{yun2024project}} & \begum{\{Python, Java, Go,} & \begum{1,500 pairs} \\
    & \begum{JS, PHP, Ruby\}} & \\
    & $\rightarrow$ English & \\

    \addlinespace[0.5ex]
    \bottomrule
  \end{tabular}
\end{table}

\begin{table}[tb]
  \caption{Overview of SQL-related Benchmarks.}
  \label{ref:table_sql}
  \centering
  \begin{tabular}{llc}
    \toprule
    \textbf{Name} & \textbf{Language(s)} & \textbf{\# Problems} \\
    \addlinespace[0.5ex]
    \midrule
    BIRD ~\cite{jinyang_li_can_2023} & English $\rightarrow$ SQL & 12,751 \\
    KaggleDBQA ~\cite{chia-hsuan_lee_kaggledbqa_2021} & English  & 272, paired with  \\
    & $\rightarrow$ SQL & golden solutions \\
    StacQc ~\cite{zhiliang_yao_staqc_2018} & English  &\{147,546 / 119,519\} \\
     & $\rightarrow$ \{Python/SQL\} & question-answer pairs\\
    Spider(V2\footnote{see \autoref{ref:table_crowd}})~\cite{lei2024spider20evaluatinglanguage} & English $\rightarrow$ SQL & 632 queries \\
    Spider-Syn~\cite{gan_towards_2021} & English $\rightarrow$ SQL & (7000 / 1034) \\
    Spider-Real~\cite{deng_structure-grounded_2021} & English $\rightarrow$ SQL & 508 \\ 
    Spider-DK~\cite{gan_exploring_2021} & English $\rightarrow$ SQL & 535 pairs \\
    Spider-CN~\cite{min_pilot_2019} & Chinese $\rightarrow$ SQL & 9691 queries \\
    SParC~\cite{yu_sparc_2019} & English $\rightarrow$ SQL & 4,298 question sequences \\
    Lyra~\cite{liang_lyra_2022} & \{English, Chinese\} & 2000 \\
    & $\rightarrow$ \{python, SQL\} & \\
    DuSQL~\cite{wang_dusql_2020} & Chinese & 23,797 \\
    & $\rightarrow$ SQL & question/SQL pairs\\
    CoSQL~\cite{yu_cosql_2019} & English $\rightarrow$ SQL & 3,007 Question Sequences \\
    \begum{SynSQL-2.5M~\cite{li2025omnisql}} & \begum{English $\rightarrow$ SQL} & \begum{2,544,390} \\
    \begum{PAUQ~\cite{bakshandaeva-etal-2022-pauq}} & \begum{Russian $\rightarrow$ SQL} & \begum{9,691} \\
    \begum{Ar-Spider~\cite{arspider}} & \begum{Arabic $\rightarrow$ SQL} & \begum{9,691} \\
    \begum{Tur2SQL~\cite{tur2sql}} & \begum{Turkish $\rightarrow$ SQL} & \begum{10,809 question/SQL pairs} \\
    
    \addlinespace[0.5ex]
    \bottomrule
  \end{tabular}
\end{table}

\normalsize


 \autoref{tab:table_cp_cc_ce} features benchmarks with competitive programming as their root, i.e., those used for understanding code complexity and efficiency. 
 \autoref{tab:table_ds} features a set of benchmarks specifically designed to evaluate the performance of models on Data Science-related tasks along with some other domain-specific SE tasks. A notable example in this table is the Bio-Coder series of benchmarks specifically designed for bioinformatics tasks. 
 To assess the mathematical reasoning capabilities of AI4SE models, see \autoref{ref:table_math}. Besides numbers and code, natural language is also a key component in AI4SE. From supporting instruction-tuned AI4SE models, which align more with the human brain~\cite{instructiontuning-2023}, that aim to accomplish question and answering (QA) similar to the widely recognized platform StackOverflow, to summarizing code and generating tags, \autoref{ref:table_nl} and \autoref{ref:table_nl2} feature AI4SE benchmarks including natural language: text-to-code, code-to-text and text-to-text (code related). We have intentionally isolated all SQL-related benchmarks due to their abundance and to facilitate locating the correct benchmark; presented in \autoref{ref:table_sql}.

\normalsize
\footnotesize

\begin{table*}[tb]
  \caption{Overview of Selected Real-to-Life SE Benchmarks. (Note: X/Y/Z denotes Train/Dev/Test)}
  \label{ref:table_general_scope}
  \centering
  \begin{tabular}{lllc}
    \toprule
    \textbf{Category} & \textbf{Benchmark} & \textbf{Language(s)} & \textbf{\# Problems} \\
    \addlinespace[0.5ex]
    \midrule
    
    \multirow{9}{*}{\shortstack[l]{Software Development \\ \& Agent Benchmarks}}
    & DevBench~\cite{devbench-2024} & Python, C/C++, Java, JS & 22 repositories \\
    & DevEval~\cite{DevEval-2024} & Python & 1,874 \\
    & CoderUJB~\cite{CoderUJB-2024} & Java & 2,239 \\
    & CODAL~\cite{m_weyssow_codeultrafeedback_2024} & Python & 500 \\
    & ToolQA~\cite{zhuang_toolqa_2023} & Python, Math, English & 800(Easy)/730(Hard) \\
    & MIT~\cite{wang_mint_2024} & Python, English & 586 Problems \\
    & SAFIM~\cite{gong_evaluation_2024} & Python, Java, C++, C\# & 17720 \\
    & AgentBench~\cite{liu_agentbench_2023} & N/A & 1360 prompts \\
    & \begum{CSR-Bench~\cite{xiao2025csr}} & \begum{Python} & \begum{100 repositories}\\
    
    \midrule
    \multirow{7}{*}{Class Level}
    & ClassEval~\cite{ClassEval-2023} & Python & 100 \\
    & CONCODE~\cite{Concode-2018} & English, Java & 104,000 \\
    & BigCodeBench~\cite{BigCodeBench-2024} & Python & 1,140 \\
    & \begum{OOP-Bench~\cite{wang2024oop}} & \begum{Python} & \begum{431} \\
    & \begum{CodeSense~\cite{roy2025codesenserealworldbenchmarkdataset}} & \begum{Python, C, Java} & \begum{2125 (Python), 876 (C)} \\
    & & & \begum{875 (Java} \\
    & \begum{ClassEval-T~\cite{classevalt}} & \begum{Java, C++} & \begum{1,243)} \\
    
    \midrule
    \multirow{20}{*}{Project \& Cross-file}
    & SWE-bench~\cite{SWE-bench-2023} & Python & 19,008 (Train), 225 (Dev),\\
    & & & 2,294 (Test) \\
    & & & 144 (Small) \\
    & CrossCodeEval~\cite{CrossCodeEval-2023} & C\#, TS, Java, Python & 2,665 (Python), 2,139 (Java),\\
    & & & 3,356 (TS), 1,768 (C\#) \\
    & CoderEval~\cite{CoderEval-2023} & Java, Python & 230 \\
    & DotPrompts~\cite{agrawal_guiding_2023} & Java & 105538 problems \\
    & & & (1420 methods) \\
    & BigCloneBench~\cite{svajlenko_evaluating_2015} & Java & 25,000 Java Systms \\
    & DI-Bench~\cite{zhang2025dibenchbenchmarkinglargelanguage} & Python, C\#, Rust, JS & 581 repositories \\
    & & & (w/ dependencies) \\
    & DyPyBench~\cite{Bouzenia_2024} & Python & 50 repositories \\
    & \begum{Multi-SWE} & \begum{Java, TS, JS} & \begum{500 (Python), 128 (Java),} \\
    & \begum{-bench~\cite{zan2025multi}}& \begum{Go, Rust, C, C++} & \begum{224 (TS), 356 (JS),} \\
    & & & \begum{428 (Go), 239 (Rust)} \\
    & & & \begum{128 (C), 129 (C++)} \\
    & \begum{KernelBench~\cite{ouyang2025kernelbench}} & \begum{Python} & \begum{250}\\
    & \begum{CodeMEnv~\cite{cheng2025codemenv}} & \begum{Python, Java} & \begum{587 (Python), 335 (Java)} \\
    & \begum{CodeEditorBench~\cite{guo2025codeeditorbench}} & \begum{Python, Java, C++} & \begum{7,961} \\
    & \begum{ProjectEval~\cite{liu2025projecteval}} & \begum{Python} & \begum{284} \\
    
    \addlinespace[0.5ex]
    \bottomrule
  \end{tabular}
\end{table*}

\begin{table}
  \caption{Overview of Selected Real-to-Life SE Benchmarks. This table contains the Repository-Level benchmarks. (Continued - Note: X/Y/Z denotes Train/Dev/Test)}
  \label{ref:table_general_scope2}
  \centering
  \begin{tabular}{llc}
    \toprule
    \textbf{Benchmark} & \textbf{Language(s)} & \textbf{\# Problems} \\
    \addlinespace[0.5ex]    
    \midrule
    RepoBench~\cite{RepoBench-2023} & Python, Java & Cross-file: 8,033\\
     & & In-file: 7,910 \\
    
    RepoEval~\cite{RepoEval-2023} & Python & 1,600 (line),  \\
    & & 1,600 (API) \\
    & & 373 (function) \\ 
    EvoCodeBench~\cite{EvoCodeBench-2024} & Python & 275 \\
    SketchEval~\cite{daoguang_zan_codes_2024} & Python & 19 repositories \\
    & & (5 easy, 8 medium \\
    & & 6 hard) \\
    Stack Repo~\cite{daoguang_zan_codes_2024} & Python & (435,890 / 220,615 \\
    & & / 159,822) answer pairs\\
    ML-BENCH~\cite{tang_ml-bench_2024} & Python \& Bash & 9641 problems \\
    CodeGen4Libs~\cite{liu_codegen4libs_2023} & Java & 403,780 prompts \\
    \begum{SWE-rebench~\cite{badertdinov2025swe}} & \begum{Python} & \begum{294 tasks, 169 repositories}\\ 
    
    \begum{SWE-Polybench~\cite{rashid2025swe}} & Java & 165 (Java) \\
    & JS & 1017 (JS), \\
    & TS& \begum{729 (TS)}\\
    & \begum{Python}& \begum{729 (199 (Python)}\\
    
    \begum{HumanEvo~\cite{zheng2025humanevo}} & \begum{Python, Java} & \begum{200 (Python), 200 (Java)}\\
    \begum{REPOCOD~\cite{liang2024can}} & \begum{Python} & \begum{980 functions (11 projects)} \\
    \begum{FEA-Bench~\cite{li2025fea}} & \begum{Python} & \begum{83 repositories} \\
    \begum{JavaBench~\cite{cao2024javabench}} & \begum{Java} & \begum{389 methods (106 classes)} \\
    \begum{SWE-bench live~\cite{zhang2025swe}} & \begum{Java} & \begum{1,319 tasks (93 repositories)} \\
    \begum{SWE-Lancer~\cite{miserendino2025swe}} & \begum{N/A} & \begum{1,488 tasks)} \\

    \addlinespace[0.5ex]
    \bottomrule
  \end{tabular}
\end{table}

\normalsize

\footnotesize

\begin{table*}[tb]
  \caption{Overview of Selected API and Retrieval Benchmarks by Category.}
  \label{ref:table_api_retrieval}
  \centering
  \begin{tabular}{lllc}
    \toprule
    \textbf{Category} & \textbf{Benchmark} & \textbf{Sources/API(s)} & \textbf{\# Problems} \\
    \addlinespace[0.5ex]
    \midrule
    
    \multirow{7}{*}{API Prediction}
    & RestBench~\cite{RestBench-RestGPT-2023} & Spotify, TMDB & 57, 100 \\
    & APIBench-Q~\cite{APIBENCH-Q-2021} & StackOverflow, Tutorial Websites & 6,563 (Java), \\ 
    & & & 4,309 (Python) \\
    & BIKER~\cite{BIKER-Dataset-2018} & StackOverflow & 33,000 \\
    & Gorilla APIBench~\cite{Gorilla-APIBench-APIZoo-2023} & HuggingFace, TensorHub, TorchHub & 925, 696, 94 \\
    & Gorilla APIZoo~\cite{Gorilla-APIBench-APIZoo-2023} & Open submissions & -- \\ 
    & & (Google, YouTube, Zoom, etc.) &  \\
    
    \midrule
    \multirow{6}{*}{Retrieval \& Planning}
    & API-Bank~\cite{API-Bank-2023} & 73 commonly used APIs & 753 \\
    & \multirow{2}{*}{CodeRAG-Bench~\cite{CodeRAG-Bench-2024}} & Competition solutions, tutorials, & \multirow{2}{*}{25,859} \\
    & & documentation, StackOverflow, GitHub & \\
    & Search4Code~\cite{rao_search4code_2021} & Bing & 6596(java)/4974(c\#)\\
    & CoIR~\cite{li_coir_2024} & GitHub, StackOverflow, and & 2.38M (corpus)\\
    & & Various Benchmarks & 3.37(queries) \\

    \midrule
    \multirow{1}{*}{Memorization}
    & SATML-ext~\cite{al-kaswan_traces_2024} & GitHub & 1,000 samples \\

    \midrule
    \multirow{4}{*}{\begum{API Misuse Detection}}
    & \begum{ExampleCheck~\cite{zhang2018code}} & \begum{StackOverflow} & \begum{100 (Java)} \\
    & \begum{ROBUSTAPI~\cite{zhong2024can}} & \begum{StackOverflow} & \begum{1208 (18 Java APIs)} \\
    & \begum{APIMU4C~\cite{gu2019empirical}} & \begum{Juliet Test Suite, ITC, } & \begum{2272 (C)} \\
    & & \begum{OpenSSL, Curl, Httpd} & \\

    \addlinespace[0.5ex]
    \bottomrule
  \end{tabular}
\end{table*}

\normalsize

\footnotesize

\begin{table*}[tb]
  \caption{Overview of AI4SE Benchmarks Related to Pseudocode.}
  \label{ref:table_pseudo}
  \centering
  \begin{tabular}{llllc}
    \toprule
    \textbf{Category} & \textbf{Benchmark} & \textbf{Language(s)} & \textbf{\# Problems} & \textbf{Crowdsourced} \\
    \midrule
    \multirow{3}{*}{Pseudocode to Code} 
    & SPoC~\cite{SPoC-2019} & C++ & 18,356 & Yes \\
    & NAPS~\cite{NAPS-2018} & Java/UAST & 17,477 & No \\
    & \begum{PseudoEval~\cite{wu2025isolatinglanguagecodingproblemsolvingbenchmarking}} & \begum{Python, C++, Rust} & \begum{1,060} & \begum{No} \\
    \midrule
    \multirow{2}{*}{Code to Pseudocode} 
    & \multirow{2}{*}{Django~\cite{Django-2015}} & Python, English & 18,805 (Train), 1,000 (Dev), & \multirow{2}{*}{No} \\
    &  & \& Japanese & 1,805 (Test) & \\
    \bottomrule
  \end{tabular}
\end{table*}

\normalsize

\footnotesize
\begin{table}[tb]
  \caption{Overview of Selected Crowd-sourced Benchmarks (NL denotes Natural Language).}
  \label{ref:table_crowd}
  \centering
  \begin{tabular}{lp{20mm}lp{20mm}}
    \toprule
    \textbf{Benchmark} & \textbf{Language(s)} & \textbf{\# Problems} & \textbf{Source} \\
    \midrule
    WikiSQL~\cite{WikiSQL-2017} & NL$\rightarrow$SQL query & 80,654 & Amazon MTurk (2017) \\
    
    Spider~\cite{Spider-2018} & NL$\rightarrow$SQL query & 10,181 & 11 Yale students (2018) \\
    
    NL2Bash~\cite{xi_victoria_lin_nl2bash_2018} & NL$\rightarrow$Bash & 9,305 & Upwork (2018) \\
    
    \multirow{2}{*}{NAPS~\cite{NAPS-2018}} & \multirow{2}{*}{\makecell[l]{Java/UAST$\rightarrow$ \\Pseudocode}} & \multirow{2}{*}{17,477} & Self-hosted crowdsourcing, \\
    & & & programming community (2018) \\
    
    SPoC~\cite{SPoC-2019} & C++ & 18,356 & programming websites (2019) \\
    
    MBPP~\cite{MBPP-MathQA-2021} & Python & 974 & Google Research, \\
    & & & internal crowdworkers (2021) \\
    \bottomrule
  \end{tabular}
\end{table}
\normalsize
\normalsize
While translating natural language is more trivial nowadays, translating code remains challenging due to various reasons (e.g. versioning, semantics, dependencies). With the lack of diversity in language support for AI4SE benchmarks and also benefiting numerous other SE tasks, \autoref{ref:table_trans} features an overview of resources that can support the ongoing development of code translation.

\footnotesize
\begin{table*}[tb]
  \caption{Overview of Programming Language Translation Benchmarks (Note: X/Y/Z denotes Train/Dev/Test).}
  \label{ref:table_trans}
  \centering
  \begin{tabular}{lllc}
    \toprule
    \textbf{Category} & \textbf{Name} & \textbf{Language(s)} & \textbf{\# Samples} \\
    \addlinespace[0.5ex]
    \midrule
    \multirow{13}{*}{Programming Languages}
    & CodeTrans~\cite{CodeXGLUE-2021} & C\#, Java & 11,800 \\
    & TransCoder-ST~\cite{TransCoderST-2022} & C++, Java, Python & 437,030 \\
    & CoST~\cite{CoST-2022} & 7 programming languages & 16,738 \\
    & AVATAR~\cite{AVATAR-2023} & Java, Python & 7,133 / 476 / 1,906 \\
    & Multilingual-Trans~\cite{CodeTransOcean-MultilingualTrans-NicheTrans-LLMTrans-DLTrans-2023} & 8 programming languages & 30,419 total \\
    & NicheTrans~\cite{CodeTransOcean-MultilingualTrans-NicheTrans-LLMTrans-DLTrans-2023} & Various niche languages & 236,468 total \\
    & LLMTrans~\cite{CodeTransOcean-MultilingualTrans-NicheTrans-LLMTrans-DLTrans-2023} & 8 programming languages & 350 \\
    & G-TransEva~\cite{jiao_evaluation_2023} & 5 programming languages & 400 total \\
    & CODEDITOR~\cite{zhang_multilingual_2023} & C\# \& Java & 6613 \\
    & \begum{RustRepoTrans~\cite{ou2024repository}} & \begum{C++, Java, Python $\rightarrow$ Rust} & \begum{375} \\
    & \begum{AVATAR-TC~\cite{jana2024cotran}} & \begum{Java, Python} & \begum{55,179 / 443 / 1,746} \\
    & \begum{RustRepoTrans~\cite{ou2024repository}} & \begum{C++, Java, Python $\rightarrow$ Rust} & \begum{375} \\
    & \begum{RepoTransBench~\cite{wang2024repotransbench}} & \begum{Python $\rightarrow$ Java} & \begum{100 repositories} \\
    
    \midrule
    Libraries & DLTrans~\cite{CodeTransOcean-MultilingualTrans-NicheTrans-LLMTrans-DLTrans-2023} & PyTorch, TensorFlow, \\& & MXNet, Paddle & 408 total \\
    \midrule
    Intermediate Representation & SLTrans~\cite{paul_ircoder_2024} & 14 Languages $\rightarrow$ LLVM-IR & 4M \\
    \midrule
    \multirow{2}{*}{\shortstack[l]{Language Conversion \\ Frameworks}}
    & MultiPL-E~\cite{MultiPL-E-2022} & 19 programming languages & - \\
    & MultiEval~\cite{MultiEval-HumanEval-MBXP-MathQA-X-2022} & 13 programming languages & - \\
    \addlinespace[0.5ex]
    \bottomrule
  \end{tabular}
\end{table*}

\normalsize

However, the workflow of a developer is not merely an exercise of writing snippets of code for given descriptions, but rather having a general overview of a project and how one can implement functionality such that it fits well into a collective code base. For this, \autoref{ref:table_general_scope} and \autoref{ref:table_general_scope2} provide a collection of benchmarks that examine the capabilities of models to generate code on a larger scale. 

The utilization of APIs plays a significant role in AI4SE benchmarks, specifically for models with Retrieval Augmented Generation (RAG) capabilities. In \autoref{ref:table_api_retrieval}, prominent benchmarks focusing on leveraging the power of APIs are denoted. Furthermore, \autoref{ref:table_pseudo} lists benchmarks related to pseudocode, followed by an overview of notable crowd-sourced AI4SE resources in \autoref{ref:table_crowd}.

\footnotesize

\begin{table}[tb]
  \caption{Overview of Automated Program Repair, Fault Localization, and Vulnerability Detection Benchmarks.}
  \label{ref:apr_vul}
  \centering
  \begin{tabular}{lllc}
    \toprule
    & \textbf{Benchmark} & \textbf{Language(s)} & \textbf{\#Samples} \\
    \midrule
    
    \multirow{19}{*}{\rotatebox[origin=c]{90}{\shortstack{Automated Program Repair \\ \& Fault Localization}}}
    & Defects4J~\cite{Defects4J-2014} & Java & 835 \\
    & GitBug-Java~\cite{GitBug-Java-2024} & Java & 199 \\
    & EvalGPTFix ~\cite{quanjun_zhang_critical_2023} & Java & 4530 \\
    & TutorCode ~\cite{boyang_yang_cref_2024} & C++ & 1239 \\
    & GHRB ~\cite{jae_yong_lee_github_2023} & Java & 107 \\
    & IntroClass ~\cite{claire_le_goues_manybugs_2015} & C & 998 \\
    & ManyBugs ~\cite{claire_le_goues_manybugs_2015} & 7 Languages & 185 \\
    & DebugBench~\cite{runchu_tian_debugbench_2024} & C++, Java & 1,438 \& 1,401 \\
    & & Python & \& 1,414 \\
    & QuixBugs~\cite{derrick_lin_quixbugs_2017} & Java & 40 (locations of bugs) \\
    & RES-Q~\cite{beck_labash_res-q_2024} & Python, JS & 100 hand-crafted \\
    & & & questions + tests \\
    & StudentEval~\cite{hannah_mclean_babe_studenteval_2023} & Python & 1,749 buggy programs \\
    & & & (48 * 3 tests) \\
    & Re-Factory~\cite{Hu_refactory_2019} & Python & 1783(buggy)\\
    & & & 2442(correct)\\
    & ConDefects~\cite{wu_condefects_2023} & Python & 526(Python) \\
    & & Java & Java(477) \\
    & Cerberus~\cite{shariffdeen_cerberus_2023} & C, C++, Java & 2242 (across 4 tasks) \\
    & \begum{RepairBench~\cite{silva2025repairbench}} & \begum{Java} & \begum{574} \\
    & \begum{MaRV~\cite{nunes2025marv}} & \begum{Java} & \begum{693} \\
    & \begum{BugsinPy~\cite{widyasari2020bugsinpy}} & \begum{Python} & \begum{493 bugs} \\

    \midrule
    \multirow{17}{*}{\rotatebox[origin=c]{90}{\shortstack{Vulnerability Detection}}}
    & CVEFixes~\cite{guru_prasad_bhandari_cvefixes_2021} & Various & 5,365  \\
    & LLMSecEval~\cite{catherine_tony_llmseceval_2023} & C & 150 (on 25  \\
    &  &  & vulnerabilities)  \\
    & SecurityEval~\cite{mohammed_latif_siddiq_securityeval_2022} & 6 languages & 130 (on 75   \\
    &  &  & vulnerabilities)  \\
    & Vul4J~\cite{bui_vul4j_2022} & Java & 79 vulnerabilities \\
    & FormAI~\cite{tihanyi_formai_2023} & C & 112k instances \\
    & VJBbench~\cite{wu_how_2023} & Java & 42 vulnerabilities \\
    & SmartBugs~\cite{durieux_empirical_2020} & Solidity & 69 Vulnerable \\
    & &  &  Smart Contracts \\
    & Devign~\cite{zhou_devign_2019} & C & 4 large  \\
    & &  & Software Repos \\
    & D2A~\cite{zheng_d2a_2021} & C/C++ & 6 OSS Programs \\
    & BigVul~\cite{10.1145/3379597.3387501} & C/C++ & 348 Projects \\
    & SARD~\tablefootnote{\href{https://samate.nist.gov/SARD/}{https://samate.nist.gov/SARD/}} & Java, C, C++ & 32k\footnote{As of 4th Feb 2025} \\
    & & C\#, PHP & \\
    & Juliet 1.3~\tablefootnote{https://samate.nist.gov/SARD/test-suites/112} & C/C++ & 64k\tablefootnote{As of 4th Feb 2025} \\
    & NVD~\tablefootnote{https://nvd.nist.gov/developers/data-sources} & Various & 265k\tablefootnote{As of 4th Feb 2025} \\
    & \begum{ARVO~\cite{mei2024arvo}} & \begum{C, C++} & \begum{1,001 vulnerabilities} \\
    & \begum{VADER~\cite{liu2025vader}} & \begum{15 languages} & \begum{174 vulnerabilities} \\
    & \begum{ManyVuls4J~\cite{lin2024there}} & \begum{Java} & \begum{103 vulnerabilities} \\

    \midrule
    \multirow{11}{*}{\rotatebox[origin=c]{90}{\shortstack{Software Testing}}}
    & CoverageEval~\cite{michele_tufano_predicting_2023} & Python & 1160 \\ 
    & ATLAS~\cite{watson_learning_2020} & Java & 9,275 projects \\
    & HITS~\cite{wang_hits_2024} & Java & 10 projects \\
    & MeMo~\cite{bareis_code_2022} & Java & 9 projects \\
    & MLAPIs~\cite{wan_automated_2022} & Python & 63 applications \\
    & \begum{CoderUJB~\cite{zeng2024coderujb}} & \begum{Java} & \begum{2,239} \\
    & \begum{TestBench~\cite{zhang2024testbench}} & \begum{Java} & \begum{108} \\
    & \begum{TestEval~\cite{wang2024testeval}} & \begum{Python} & \begum{210} \\
    & \begum{TARBENCH~\cite{yaraghi2025automated}} & \begum{Java} & \begum{45,373 (59 projects)} \\
    & \begum{ProjectTest~\cite{wang2025projecttest}} & \begum{Python, Java, } & \begum{20 per language} \\
     & & \begum{JS} &  \\
    & \begum{CLOVER~\cite{xu2025clovertestcasegeneration}} & \begum{Python} & \begum{845} \\
    \bottomrule
  \end{tabular}
\end{table}

\begin{table*}[tb]
  \caption{Overview of Selected SE-Workflow Benchmarks.}
  \label{ref:se_related}
  \centering
  \begin{tabular}{lllc}
    \toprule
    \textbf{Category} & \textbf{Benchmark} & \textbf{Language(s)} & \textbf{No. of Samples} \\
    \midrule
    \multirow{8}{*}{\shortstack{Code Synthesis \\ \& Understanding}}
    & Methods2Test~\cite{Methods2Test-2020} & Java & 780,944 \\
    & CRUXEval~\cite{CRUXEval-2024} & Python & 800 \\
    & CRQBench~\cite{elizabeth_dinella_crqbench_2024} & C++ & 100 \\
    & CriticBench~\cite{lin_criticbench_2024} & Python & 3,825(across 5 tasks)\\
    & CodeScope~\cite{yan_codescope_2024} & 8 PLs & 13,390 (across 8 tasks)\\
    & \begum{CodeCriticBench~\cite{zhang2025codecriticbench}} & \begum{Various} & \begum{1,517 (Easy), 1,084 (Medium),} \\  & & & \begum{1,699 (Hard)} \\
    & \begum{CRUXEval-X~\cite{xu2025cruxevalxbenchmarkmultilingualcode}} & \begum{19 PLs} & \begum{19K} \\
    
    \midrule
    
    \multirow{1}{*}{Merge Conflict Repair}
    & ConflictBench~\cite{ConflictBench-2024} & Java & 180 \\
    
    \midrule

    \multirow{2}{*}{Type Inference}
    & TypeEvalPy ~\cite{ashwin_prasad_shivarpatna_venkatesh_typeevalpy_2023} & Python & 845 (annotated labels) \\   
    & TypeEvalPy AutoGen ~\cite{ashwin_prasad_shivarpatna_venkatesh_typeevalpy_2023} & Python & 78373 (annotated labels) \\   
    \midrule
    
    \multirow{7}{*}{\shortstack{Automatic Code \\ Quality Review}}
    & CodeReview ~\cite{zhiyu_li_automating_2022} & 8 languages & 7.9M pull requests \\
    & Software Maintainability ~\cite{markus_schnappinger_defining_2020} & Java & 519 projects \\ & & &(evaluations of quality) \\
    & \begum{BenMark~\cite{wang2025deep}} & \begum{Java} & \begum{1,299,186 methods} \\
    & \begum{CodeReviewer~\cite{li2022automating}} & \begum{9 languages} & \begum{13,100} \\
    & \begum{CodeReview-New~\cite{guo2024exploring}} & \begum{16 languages} & \begum{14,600} \\
    & \begum{CodeReviewQA~\cite{lin2025codereviewqa}} & \begum{9 languages} & \begum{900} \\
    
    \midrule
    \multirow{5}{*}{Hallucination Detection}
    & HALLUCODE~\cite{fang_liu_exploring_2024} & Python & 5,663 \\
    & \begum{CodeHaluEval~\cite{tian2025codehalu}} & \begum{Python} & \begum{699} \\
    & \begum{Collu-Bench~\cite{jiang2024collu}} & \begum{N/A} & \begum{13,234} \\
    & \begum{LMDefects~\cite{fan2023automated}} & \begum{Java} & \begum{60 (Easy), 53 (Medium)} \\
    & \begum{Codemirage~\cite{agarwal2024codemirage}} & \begum{Python} & \begum{1,137} \\
    \bottomrule
  \end{tabular}
\end{table*}

\normalsize

 With AI4SE models mainly being utilized for program synthesis, it remains relatively questionable how effective these models are in generating tests and repairing bugs, as it is unclear whether these models \textit{truly} understand code. For example, Siddiq et al.~\cite{siddiq2024using} observed Codex~\cite{HumanEval-2021} being able to get above 80\% coverage for HumanEval~\cite{HumanEval-2021}, yet many test smells were discovered and for another dataset, no higher than 2\% coverage was attained. This reveals the importance of benchmarking AI4SE models' capabilities in test generation, bug repair, and understanding. In \autoref{ref:apr_vul}, several benchmarks are listed that make an effort to assess the aforementioned. \roham{Additionally, \autoref{ref:se_related} table features benchmarks that have been designed to evaluate a model's capabilities in dealing with everyday tasks of a software engineer (e.g., merge-conflict repair, Code Reviews, etc.).}

\footnotesize
\begin{table*}[tb]
  \caption{Overview of Multi-Category Benchmarks, Covering Various Tasks.}
  \label{ref:table_multi_categoy}
  \centering
  \begin{tabular}{lllc}
    \toprule
    \textbf{Name} & \textbf{Language(s)} & \textbf{Tasks} & \textbf{Information} \\
    \addlinespace[0.5ex]
    \midrule

    Big-Bench~\cite{BIG-Bench-2022} & \makecell[l]{Python, Numeric, \\ JSON, English}& \makecell[l]{Functions over numbers,\\Mathematical Reasoning, \\Text2Code, Code2Text, \\ Code explanation, Debugging, \\ Turing Complete \\ Concept Learning, \\ \href{https://github.com/google/BIG-bench/blob/main/bigbench/benchmark_tasks/README.md}{amongst other tasks}} & \makecell[l]{250, \\ several per category,\\ 42, 60, 66, \\ 34, 6,390} \\
    
    \midrule

    XLCoST~\cite{XLCoST-2022} & \makecell[l]{C, C++, C\#, Java, JS, \\ Kotlin, PHP, Python, Ruby, \\ Rust} & \makecell[l]{Text2Code \\ (program synthesis, code search), \\ Code Summarization, \\ Code Translation} & \makecell[l]{567K \\ (509k, 58k), \\ 567K, 122K} \\
    
    \midrule

    CrossCodeBench~\cite{changan_niu_crosscodebench_2023} & \makecell[l]{Java, C\#, Python, C++, \\ JS, PHP, Go, Ruby, \\ TS, C, Bash, Shell} & \makecell[l]{Classification, In-Filling, \\\ Translation, Generation, \\ Summarization, Type Prediction, \\ Question Answering} & \makecell[l]{6.6M, 13.4M, 2.4M, \\ 19.5M, 11.2M,\\ 773K, \\ 190K} \\

    \midrule
    Long Code Arena~\cite{long-code-arena-2024} & \makecell[l]{English, Python, Java, Kotlin} & \makecell[l]{Commit Message Generation, \\ Module Summarization, \\ Library-Based Code Generation, \\ Project-Level Code Completion, \\ Bug Localization, CI Builds Repair} & \makecell[l]{163, \\ 216, \\ 150, \\ 908 (varying sizes), \\ 14.96K, 78} \\

    \midrule

    \makecell[l]{CodeXGLUE ~\cite{long-code-arena-2024}\\MicrosoftDocs\tablefootnote{\url{https://github.com/MicrosoftDocs/}}\\CodeSearchNet~\cite{CodeSearchNet-Challenge-2019}} & \makecell[l]{English, Chinese, \\ Norwegian, Danish, Latvian \\ Go, Java, JS, \\PHP, Python, Ruby} & \makecell[l]{Code Documentation Translation,\\ Code Documentation \\ (Code Summarization, \\ Comment Generation)} & \makecell[l]{(CN: 52K, NOR: 26K, \\ DK: 45K, LT: 21K), \\ 621870} \\ 

    \midrule

    \begum{DomainEval~\cite{zhu2025domaineval}} & 
    \begum{Python} & 
    \begum{\makecell[l]{Computation, Network, \\ Basic operation, System, \\ Visualization, Cryptography}} & 
    \begum{5,892 cases total} \\

    \midrule
    
    \begum{CodeApex~\cite{liu2022code}} & 
    \begum{C++, English, Chinese} & 
    \begum{\makecell[l]{Programming comprehension \\ Code generation, Code correction}} & 
    \begum{250, 476, 1330} \\
        
    \bottomrule
  \end{tabular}
\end{table*}

\normalsize

Additionally, we collect a set of generic benchmarks in \autoref{ref:table_multi_categoy}. These benchmarks are not only single-task benchmarks like others previously seen in this section but are rather a collection of tasks spanning a wide range of languages and task types. A notable, and widely known benchmark in this category is the BigBench Benchmark~\cite{BIG-Bench-2022} which consists of 167 tasks (not all relevant to AI4SE).

\footnotesize
\begin{table}[tb]
\color{navy}
  \caption{Overview of Selected Benchmarks for Log Statement Generation and Parsing.}
  \label{ref:table_log}
  \centering
  \begin{tabular}{p{40mm}llp{20mm}}
    \toprule
    \textbf{Category} & \textbf{Benchmark} & \textbf{Language(s)} & \textbf{\#Problems} \\
    \addlinespace[0.5ex]
    \midrule

    \multirow{4}{*}{\rotatebox[origin=c]{0}{\shortstack{Log Statement\\Generation}}}
    & LANCE~\cite{mastropaolo2022using} & Java & 76,421 \\
    & LogBench~\cite{li2023exploring} & Java & 6,849 \\
    & SCLogger~\cite{li2024go} & Java & 31,170 \\
    & AL-Bench~\cite{tan2025albenchbenchmarkautomaticlogging} & Java & 39,600 \\
    
    \midrule

    \multirow{5}{*}{\rotatebox[origin=c]{0}{\shortstack{Log Parsing}}}
    & LogBase~\cite{zhang2025logbase} & Various & 85,300 \\
    & LogHub~\cite{zhu2023loghub} & Various & 32,000 \\
    & LogHub-2.0~\cite{jiang2024large} & Various & 3.6M \\
    & LogPM~\cite{logpm} & Various & 10,821,589 \\
    & LogEval~\cite{cui2024logevalcomprehensivebenchmarksuite} & Eng., Chin. & 4K logs (4 tasks) \\

    \addlinespace[0.5ex]
    \bottomrule
  \end{tabular}
\end{table}

\normalsize

\rohamrev{Finally, we have identified a rather large count of papers that have to do with the matter of Log Parsing and Log Statement generation. we have presented these benchmarks in \autoref{ref:table_log}.}

\begum{\subsection{Limitations of Existing Benchmarks}}
\begum{\begum{
\begin{table*}[tb]
\color{navy}
  \caption{Common limitations in AI4SE benchmarks}
  \label{tab:benchmark-limitations}
  \centering
  \begin{tabular}{p{3.7cm} p{7cm} p{4cm} p{1.5 cm}}
    \toprule
    \textbf{Limitation} & \textbf{Description} & \textbf{Representative Examples} &
    \textbf{Frequency Observed}
    \\
    \midrule

    Language Specific & \makecell[l]{Focuses only on one programming language} & \makecell[l]{HumanEval, MBPP, APPS} & 169 \\
    \midrule

    No Leaderboard & \makecell[l]{No official leaderboard; hinders fair and consistent\\ comparisons} & \makecell[l]{SPoC, RepoEval,\\ CONCODE} & 201 \\
    \midrule

    \makecell[l]{Poor Maintenance/Data Quality} & \makecell[l]{Dataset contains outdated problems, lacks documentation,\\ or is not updated regularly} & \makecell[l]{ToolQA, ExampleCheck} & 218\\
    \midrule

    Not Peer Reviewed & \makecell[l]{Not been published in a peer-reviewed venue} & \makecell[l]{CodeSearchNet, Methods2Test} & 79\\
    \midrule

    Infrequently Used & \makecell[l]{Rarely cited in academic work} & \makecell[l]{JavaBench, CoIR} & 199 \\
    \midrule

    Dataset Creation Method & \makecell[l]{How problems were obtained: \textit{mined} from real-world sources,\\ \textit{created}  by humans, or \textit{LLM-generated}} & \makecell[l]{AVATAR (mined),\\ Spider (created), \\ CRUXEval (generated)} & 
    \makecell[l]{Mined: 149,\\ Created: 69,\\ Generated: 46}\\
    \bottomrule
  \end{tabular}
\end{table*}
}

\normalsize
}

\begum{Across our review, we observed several trends in the existing AI4SE benchmarks. The majority are language-specific, most commonly targeting Python or Java. Although these languages are popular in both education and industry, this narrow scope potentially limits their generalization across domains and programming environments. Many benchmarks suffer from poor maintenance, with little to no updates or active support following their initial release. Official and dynamic leaderboards are often missing, which makes it difficult to fairly compare model performance across time or publications, particularly as newer models continue to emerge. Notably, benchmark popularity does not appear to strongly correlate with its maintenance quality or whether it was peer reviewed, suggesting other factors such as visibility or convenience may drive adoption.

There is also variation in how benchmarks are constructed. Manually created datasets are often significantly smaller, whereas newer benchmarks increasingly leverage LLMs to generate large volumes of tasks. Mined datasets which are mostly sourced from GitHub, can be large but are frequently under-specified, with limited transparency into the repositories used or the sampling heuristics applied.

Table~\ref{tab:benchmark-limitations} summarizes these limitations and provides representative benchmarks for each. The full list of evaluated benchmarks and their metadata can be found in the replication package of the study.}

In conclusion, based on our in-depth inspection of HumanEval and MBPP and combined with the inspection of other benchmarks in our review, we obtained an overview of the limitations in current AI4SE benchmarks which we use as a guide in shaping our methodology proposed in \autoref{sec:humanevalpro}.


\section{\searchtool - Locating AI4SE Benchmarks}
\label{sec:searchtool}
Due to the abundance of AI4SE benchmarks, identifying the most suitable one for a specific SE task can be challenging. As a result, many default to evaluating their models on popular benchmarks like HumanEval~\cite{HumanEval-2021} which has its own flaws.

To address this gap, we developed \textsc{\searchtool}~\footnote{\url{https://evalpro.online/search.html}}, a tool to systematically and semantically search the existing benchmarks and their corresponding use cases. We additionally provide an interface to visually evaluate the closeness and similarity of a group of datasets, along with capabilities to find relations between citing bodies for identifying patterns relevant to different use cases.

\rohamrev{
\subsection{Context Extraction and Visualization}
\label{sec:searchtool:context}

\subsubsection{Overview of the Semi-Automated Pipeline.}
To effectively contextualize and visualize the growing corpus of AI4SE benchmarks, we developed a semi-automated pipeline that extracts, embeds, clusters, and visualizes benchmark metadata. As illustrated in \autoref{fig:searchtool_pipeline}, the system is organized into four modular layers: data collection, metadata enrichment, clustering and labeling, and front-end interaction.

The pipeline begins with benchmark sources collected from a curated spreadsheet, user-submitted suggestions, and an automated paper discovery script using the Semantic Scholar API to make suggestions. All inputs are aggregated into a processing queue, where metadata is enriched using both automated API calls and basic scraping techniques (e.g., fetching GitHub or Hugging Face README files). These metadata are transformed into dense vector representations using OpenAI's \texttt{ text embedding-3-small} model.

To support user navigation and cluster discovery, the high-dimensional embeddings are reduced via UMAP and grouped using HDBSCAN. A key component of the pipeline is the integration of GPT-based cluster labeling, which generates short, descriptive names for each cluster based on the titles and descriptions of the papers it contains. This step is fully automated but subject to human validation when new clusters emerge; meaning, these clusters \emph{can} be modified by humans if needed.

\subsubsection{Balancing Automation and Oversight.}
Unlike other prior approaches that rely on manual curation or entirely unsupervised grouping, our pipeline adopts a semi-automated architecture designed to scale with minimal human intervention while retaining curatorial oversight. Specifically, the system regularly ingests new suggestions, both from automated Semantic Scholar queries and from users interacting with the front-end interface, and proposes candidate entries for including in the pool of benchmarks. Although metadata extraction and embedding are fully automated, curators retain the ability to approve or reject new data sources, review GPT-generated labels, and fine-tune cluster boundaries as needed.

\subsubsection{Interactive Visualization.}
The final layer of the pipeline provides an interactive 2D interface that enables users to explore benchmarks based on their semantic similarity. The visualization supports fuzzy search, local citation graphs, and interactive paper details. Importantly, user interactions on the frontend, such as suggesting new benchmarks, are routed back into the data collection layer (cf. \autoref{fig:searchtool_pipeline}).

\subsection{Additional Features}
To improve the search and exploration experience, we incorporated several key functionalities. First, a \textbf{text-based search interface} allows users to find articles by title and abstract using \textbf{fuzzy search} via the Fuse.js library\footnote{\url{https://www.fusejs.io/}}. Users can also issue \textbf{advanced field-specific queries} (e.g., \texttt{language:python}) or \textbf{exact phrase queries} (e.g., \texttt{"code generation"}), and benefit from \textbf{autocomplete suggestions} for common terms, programming languages, tasks, and datasets.

A dynamic \textbf{2-D UMAP visualization} displays papers embedded in semantic space. Hovering reveals abstracts via a \textbf{Paper Content Tooltip}, while double-clicking redirects to the paper’s DOI page. Paper nodes are sized based on \textbf{normalized citation count}, colored by cluster, and annotated with metadata such as year, tasks, and dataset names.

Clicking a point activates the \textbf{Related Papers feature}, which shows (1) The top-5 most similar papers based on cosine similarity, (2) A \textbf{detailed paper overview} with title, abstract, authors, venue/journal, and publication type, (3) A list of \textbf{citing papers} with sortable tables and filters by year and venue, and (4) An interactive \textbf{Paper Citations Graph}, which visualizes inter-citation relationships among citing papers.

Users can navigate between \textbf{chart}, \textbf{list}, and \textbf{grid} views, paginate results, and apply \textbf{advanced filters} by programming/natural language, dataset, task, cluster, year, and citation count. A \textbf{search history} and \textbf{keyboard shortcuts} improve usability. The entire interface supports responsive interaction with detailed feedback and error handling.
}

\begin{figure}[t]
\centering
\resizebox{0.98\linewidth}{!}{%
\begin{tikzpicture}[
    every node/.style={font=\small, align=center, draw, rounded corners},
    box/.style={minimum width=2.6cm, minimum height=1cm, draw, rounded corners},
    every node/.append style={transform shape}
]

\node[box] (usersugg) at (0,0) {User Suggestions};
\node[box, left=0.35cm of usersugg] (google) {Curated\\Google Sheet};
\node[box, right=0.35cm of usersugg] (ssapi) {SS API\\Suggestions};

\node[box, above=0.45cm of usersugg] (metadata) {Metadata\\Enrichment};
\node[box, left=0.35cm of metadata] (ghhf) {GitHub / HF\\Descriptions};
\node[box, right=0.35cm of metadata] (embed) {OpenAI\\Embeddings};

\node[box, above=0.45cm of metadata] (umap) {UMAP\\Reduction};
\node[box, left=0.35cm of umap] (hdbscan) {HDBSCAN\\Clustering};
\node[box, right=0.35cm of umap] (labeling) {GPT-based\\Labeling};

\node[box, above=0.45cm of umap] (frontend) {Interactive 2D Map\\+ Fuzzy Search\\+ Related Papers};

\node[coordinate] (collect_center)  at ($(usersugg)!0.5!(metadata)$) {};
\node[coordinate] (meta_center)     at ($(metadata)!0.5!(umap)$) {};
\node[coordinate] (cluster_center)  at ($(umap)!0.5!(frontend)$) {};

\draw[->, thick] (usersugg.north) -- (collect_center);
\draw[->, thick] (metadata.north) -- (meta_center);
\draw[->, thick] (umap.north) -- (cluster_center);

\begin{pgfonlayer}{background}
    \node[
        fill=green!10, rounded corners, inner sep=6pt,
        fit=(google)(usersugg)(ssapi),
        label={[anchor=east,font=\bfseries]west:Data Collection}
    ] {};
    \node[
        fill=blue!10, rounded corners, inner sep=6pt,
        fit=(metadata)(ghhf)(embed),
        label={[anchor=east,font=\bfseries]west:Metadata \& Embeddings}
    ] {};
    \node[
        fill=red!10, rounded corners, inner sep=6pt,
        fit=(umap)(hdbscan)(labeling),
        label={[anchor=east,font=\bfseries]west:Clustering \& Labeling}
    ] {};
    \node[
        fill=yellow!20, rounded corners, inner sep=6pt,
        fit=(frontend),
        label={[anchor=east,font=\bfseries]west:Frontend Interface}
    ] {};
\end{pgfonlayer}

\end{tikzpicture}
}
\caption{Pipeline architecture of \searchtool. This includes automated metadata extraction, semantic embedding, clustering, and interactive visualization.}
\label{fig:searchtool_pipeline}
\end{figure}

\subsection{User Study}
 With all the mentioned features, we aim to create a platform that can be extended and used by both academics and practitioners alike to find the appropriate dataset/benchmark for their use case with more ease. To evaluate how effective and usable this new tool is for the end-users, we conducted a user study on \usercount people from both industry (9) and academia (13). In selecting demographics for the \searchtool user study, we aimed to assess the tool's effectiveness across a diverse group of users with varying degrees of expertise. This approach seeks to determine the tool's applicability for individuals at either end of the spectrum; whether they are beginners or seasoned experts, from academia or industry. The ultimate objective of the tool is to facilitate the selection of the appropriate benchmark, making it more accessible irrespective of prior knowledge. For this, we had each participant interact with the tool for however long they saw fit and asked them to fill out a questionnaire consisting of eleven 5-point Likert Scale questions and three open questions.

\subsubsection{Questionnaire Design}
\footnotesize
\begin{table*}
  \caption{Questionnaire Design Overview}
  \label{tab:questionnaire_design}
  \centering
  \renewcommand{\arraystretch}{1.2}
  \begin{tabular}{
    >{\raggedright\arraybackslash}m{1.7cm}  
    >{\raggedright\arraybackslash}m{8.5cm}  
    >{\centering\arraybackslash}m{2.8cm}  
    >{\centering\arraybackslash}m{2.8cm}  
  }
    \toprule
    \textbf{Section} & \textbf{Questions} & \textbf{Scale} & \textbf{Additional Info} \\
    \midrule
    
    Participant Background &
    What is your professional background? 
    What is your role? 
    How familiar are you with AI4SE benchmarks? 
    How many years of experience do you have in this field? &
    5-point Likert Scale \newline(1: Not familiar, 5: Very familiar) &
    Experience question \newline(<1, 1--3, 3--5, 5+ years) \\
    
    \midrule

    Search Functionality &
    How easy was it to navigate the interface? 
    How intuitive was the search functionality? 
    How effective was the tool in finding benchmarks? 
    Was the visual evaluation of datasets useful? &
    5-point Likert Scale \newline(1: Not useful, 5: Extremely useful) &
    N/A \\
    
    \midrule
    
    Cross-referencing Feature &
    How useful was the cross-referencing feature? \newline
    Did the tool help in understanding relationships between benchmarks? \newline
    Was the visual interface for benchmark similarity useful? &
    5-point Likert Scale, Open-ended &
    Includes qualitative feedback option \\
    
    \midrule
    
    User Experience \& Feedback &
    How would you rate the overall user experience? 
    How likely are you to use the tool in your work? 
    Did you experience any issues or challenges? 
    What other tools do you use for searching benchmarks? 
    How does this tool compare to others? 
    How well does the tool meet the needs of professionals in AI4SE? &
    Likert Scale, Open-ended &
    Open-ended questions for in-depth feedback \\
    
    \bottomrule
  \end{tabular}
\end{table*}
\normalsize
 The questionnaire designed for the study was divided into four key sections to gather feedback about the tool, namely the participant's background, the quality of the search functionality, the quality of the cross-referencing feature, and the overall user experience. \autoref{tab:questionnaire_design} is an overview of the posed questions.

\subsubsection{Results and Analysis}
 We analyzed the data collected through the questionnaire both quantitatively (Likert scale responses, scaling between 1-5) and qualitatively (open-ended questions). 
 The detailed results are provided in the replication package.~\footnote{\url{https://github.com/AISE-TUDelft/AI4SE-benchmarks}}
 Below, we present an overview.

The respondents of the questionnaire were generally familiar with AI4SE, with an average familiarity rating of 3.8. There were varying levels of experience in the field, with the range between 1-3 years being most common (eight people). 
\rohamrev{More concretely, in our pool of participants, the demographic distribution was as follows:
 \begin{enumerate}
     \item \textbf{Roles:} The participants held diverse roles, including 6 Researchers, 5 PhD Candidates, 5 Students, 4 Software/Research Engineers, and 2 Lead Researchers
     
     \item \textbf{Experience Level:} 8 participants had 1-3 years, 6 had 3-5 years, 3 had 5+ years, and 5 had less than 1 year of experience
     
     \item \textbf{Familiarity with AI4SE:} On a scale of 1 (Not familiar) to 5 (Very familiar), 8 participants rated themselves a ``5" and 7 rated themselves a ``4," indicating a strongly informed participant pool
 \end{enumerate}
 }

 In terms of search functionality, the tool scored high on usability with an average rating of 4.5, showing that users found it easy to navigate. The intuitiveness of the search interface received a solid 4.1, while its effectiveness in helping users find benchmarks was rated 4.0. The visual evaluation feature received 3.8, indicating some room for improvement in how visual elements assist in the search process.

 When evaluating the cross-referencing features, respondents found the overall usefulness to be 4.1. However, the tool's ability to help users understand connections between different benchmarks was at 3.7. The visual interface for exploring these connections was rated 3.9, suggesting that while users found it generally helpful, enhancements could improve its utility.

 Regarding the overall user experience, participants gave an average rating of 4.2, with a score of 4.0 on the likelihood of using the tool in their own research. Several issues were highlighted, particularly around the dimensionality reduction and how the scatter plot is organized and presented. Users also noted that the citation network feature becomes less effective with larger papers and called for improved clustering by topic and additional features to explain and control visualizations. 
 
The participants' responses confirmed our findings and highlighted the lack of a specific tool dedicated to locating AI4SE benchmarks. Instead, respondents commonly rely on generic platforms like Huggingface, Semantic Scholar, Google, and ConnectedPapers. When compared to these tools, \searchtool received an average score of 4.2 out of 5, with 5 indicating a much better experience. One participant mentioned using their personal network to find benchmarks, which limits broader access, further supporting the need for the proposed tool.

 The tool's ability to meet the professional needs of users was rated 4.2 which affirmed its usefulness in the AI4SE domain. However, respondents suggested several additional features that could enhance its functionality, such as pagination for citations, incorporating metadata and additional information about the papers in the search process, improved clustering and filtering options, and sorting citations based on specific criteria. Additional requests included dark mode support, better overall search functionality, and clearer explanations and control over the chart visualizations. Based on the users' feedback and components' scores, we prioritized these features and incrementally added them to the platform.

In conclusion, the tool is largely perceived as useful and user-friendly, though several areas, particularly around visualization, citation handling, and filtering options, could be improved to enhance the overall user experience and the tool’s effectiveness. Due to time constraints, we prioritized and implemented key features, leaving some for the future.

\section{\benchframework}\label{sec:humanevalpro}
In this section, we propose \textsc{\benchframework}, a detailed and peer-review-oriented approach for improving the quality of existing benchmarks. We explain our approach through a case study in which we propose \textsc{\humanevaladvanced}, a corrected foundation based on the HumanEval benchmark.

The HumanEval benchmark has long been the de facto standard for evaluating the code-generation capabilities of AI4SE models. It has recently been used to evaluate the latest and greatest LLMs from large companies such as Google Gemini \cite{Gemini-Natural2Code-2023} and OpenAI GPT4 \cite{GPT-4-2023}. Despite the great fame, however, our review in \autoref{sec:review} points towards the existence of numerous notable problems with this foundational benchmark, namely, the existence of incorrect tests, suboptimal canonical solutions, and imprecise problem definitions amongst others.

\subsection{Approach}
\begin{figure}[t]
\centering
\resizebox{\columnwidth}{!}{%
\begin{tikzpicture}[
  font=\scriptsize,
  >=Stealth,
  node distance=4mm and 4mm,
  flow/.style={-{Stealth}, line width=0.7pt},
  box/.style={draw, rounded corners=5pt, minimum height=7mm,
              text width=2.8cm, align=center, inner sep=2pt, font=\scriptsize\bfseries},
  db/.style={shape=cylinder, shape border rotate=90, draw,
             minimum height=6mm, minimum width=10mm, aspect=0.60}
]

\definecolor{laneGreen}{HTML}{D6ECE4}
\definecolor{laneBlue}{HTML}{D5E2FC}
\definecolor{boxGreen}{HTML}{4FAE9C}
\definecolor{boxBlue}{HTML}{6D93FF}
\definecolor{boxPurple}{HTML}{8C6FE9}
\definecolor{dbPurple}{HTML}{8062E3}

\node[box, fill=boxGreen!85, text=white] (full) {Full Code Review\\of HumanEval};
\node[box, fill=boxGreen!85, text=white, right=6mm of full] (std) {Standardized\\Observations};
\node[box, fill=boxPurple!85, text=white, right=6mm of std, text width=2.6cm] (check) {Check Observations};
\node[db,  fill=dbPurple!85, right=4mm of check] (dbTop) {};
\draw[flow] (full.east) -- (std.west);
\draw[flow] (std.east)  -- (check.west);
\draw[flow] (check.east)-- (dbTop.west);

\begin{pgfonlayer}{background}
  \node[fit=(full)(std), rounded corners=10pt, inner sep=4pt,
        fill=laneGreen, draw=none] (laneTop) {};
\end{pgfonlayer}

\node[anchor=north] at ($(dbTop.south)+(0,-1mm)$) {\textbf{HE Variants}};

\node[box, fill=boxBlue!85, text=white, below=12mm of full.west, anchor=west] (mod)
      {Modify HumanEval\\($\rightarrow$ HumanEvalNext)};
\node[box, fill=boxBlue!85, text=white, right=6mm of mod] (peer) {Peer-Review\\Changes};
\node[box, fill=boxBlue!85, text=white, right=6mm of peer, text width=2.5cm] (eval) {Evaluate};
\node[db,  fill=dbPurple!85, right=4mm of eval] (dbBot) {};
\draw[flow] (mod.east)  -- (peer.west);
\draw[flow] (peer.east) -- (eval.west);
\draw[flow] (eval.east) -- (dbBot.west);

\begin{pgfonlayer}{background}
  \node[fit=(mod)(peer)(eval), rounded corners=10pt, inner sep=4pt,
        fill=laneBlue, draw=none] (laneBot) {};
\end{pgfonlayer}

\node[anchor=north] at ($(dbBot.south)+(0,-1mm)$) {\textbf{AI4SE Models}};

\end{tikzpicture}%
}
\caption{\textsc{\benchframework}'s approach through a case-study of \textsc{HumanEval}.}
\label{fig:humanevalpro:approach}
\end{figure}

To improve the quality of the given benchmark, we pursue the following approach, also illustrated in \autoref{fig:humanevalpro:approach}. First, we initiate a comprehensive code review, leading to standardized observations (\autoref{sec:humanevalpro:approach:observations}). Then, we address these identified issues through a series of modifications (\autoref{sec:humanevalpro:approach:modifications}), followed by a peer review (\autoref{sec:humanevalpro:approach:peer-review}) to ensure accuracy and reliability. Finally, we experiment with the revised benchmark to evaluate and discuss the results. Below, more details are provided for specific steps in this approach.

\subsubsection{Standardized Observations in Current HumanEval Benchmarks} \label{sec:humanevalpro:approach:observations}

Upon examining and manually reconstructing canonical solutions and experimenting with various HumanEval benchmark test suites, we identified several recurring issues. The system frequently produces incorrect and sub-optimal code, as canonical solutions are inefficient and fail to address critical assumptions outlined in the problem descriptions. 

Additionally, these solutions often lack type annotations, further complicating the evaluation process. Another significant problem is the absence of quality testing. Test suites tend to overlap with example tests from the prompt, allowing incorrect canonical solutions to pass. Moreover, there are instances where the expected outputs in the test suites do not align with the canonical solutions' actual performance.

Compounding these issues is the poor quality of the problem descriptions, which contain grammatical errors, ambiguous instructions, and inconsistent formatting, particularly in the test examples. Furthermore, the system's support for language conversion frameworks, such as MultiPL-E, is inadequate. MultiPL-E, while the most comprehensive framework available, only supports equality assertions, which proves incompatible with the setup of many problems, further hindering the system's effectiveness.

\subsubsection{Modifications in \textsc{\humanevaladvanced}}
\label{sec:humanevalpro:approach:modifications}
In \textsc{\humanevaladvanced}, we address all the above issues by manually modifying all problems in the original HumanEval benchmark. 
\new{We decide to improve the original HumanEval as (1) flaws in the original version persist even in improved versions such as HumanEvalPlus, and (2) the fact that HumanEval is still widely used in new literature. While in our we have gone for the original benchmark, due to the adaptable nature of \benchframework one could opt to have an improved version of the dataset as the starting point.}
To summarize, these are the general changes made in \textsc{\humanevaladvanced} and their benefits:
In this work, several key improvements have been made to address the shortcomings of previous benchmarks. First, \textbf{all suboptimal and incorrect canonical solutions have been fixed}, which were previously missed due to insufficient testing and a lack of comprehensive quality review. Furthermore, \textbf{type annotations have been added} to all problems, offering valuable context and simplifying the translation to other programming languages. The original HumanEval benchmark only included type annotations for the first 30 problems, representing merely 18\% of the total set of 164 problems. In addition, \textbf{better support for language conversion frameworks} has been incorporated. For instance, \textsc{\humanevaladvanced} now features improved compatibility with frameworks like MultiPL-E~\cite{MultiPL-E-2022}, which supports translation to 18 additional programming languages by standardizing all tests to equality assertions where feasible. This change has reduced the number of incompatible problems by a factor of ten.

Besides these adjustments, challenging scenarios (such as negative values, zero instances, empty inputs, and non-alphanumeric symbols) are incorporated into each task to guarantee that only top-quality AI4SE models, adept at addressing diverse situations, succeed. Accordingly, \textbf{assertions are implemented} within the code wherever constraints are detailed in the problem description. This measure prevents models from ignoring crucial details, thus improving the benchmark's worth. \new{In particular, we employ specification-based testing to assess functions; using boundary analysis, we explore combinations of within, on, and outside points.} Furthermore, the \textbf{test examples in the problem descriptions have been refined}, which significantly affects model performance~\cite{MBPP-MathQA-2021}. Problems with excessive test examples now feature a reduced set, distributing the evaluation workload more fairly. Lastly, \textbf{spelling errors have been corrected}, descriptions have been consistently formatted, and \textbf{problem descriptions have been aligned with the implementations}, while still leaving room for models to demonstrate intuitive problem-solving skills expected of high-quality AI4SE models. The difficulty level has also been raised by incorporating \textbf{more edge cases and modifying various problems}, aiming to better reflect real-world challenges faced by engineers and to mitigate issues related to data leakage and saturated performance on the HumanEval leaderboards.
To illustrate the modifications of the tests in \textsc{\humanevaladvanced}, consider  \autoref{tab:humanevalpro:comparison-test-statistics}.

\begin{table}[tb]
\centering
\caption{Comparison of test statistics between HumanEval (based on \url{https://github.com/openai/human-eval/blob/master/data/HumanEval.jsonl.gz}{\texttt{human-eval-v2-20210705.json}}) and \textsc{\humanevaladvanced}.}
\label{tab:humanevalpro:comparison-test-statistics}
\begin{tabular}{|l|c|c|c|}
\hline
\textbf{Metric} & \textbf{HumanEval} & \textbf{\textsc{\humanevaladvanced}} & \textbf{$\Delta$} \\
& \textbf{(original)} & & \\
\hline
Total number of asserts & 1325 & 2551 & $\times1.92$ \\
\hline
Avg. number of asserts & 8 & 16 & $\times2$ \\
\hline
Med. number of asserts & 7 & 11 & $\times1.57$ \\
\hline
Min. number of asserts & 1 & 4 & $+3$ \\
\hline
Total with $<$ 5 asserts & 34 & 2 & -94\% \\
\hline
\end{tabular}
\end{table}

\subsubsection{Peer Review Process of \textsc{\humanevaladvanced}} \label{sec:humanevalpro:approach:peer-review}
To ensure the accuracy and reliability of \textsc{\humanevaladvanced}, an independent reviewer \new{verified} all changes. This thorough review involved verifying the clarity and completeness of the problem docstrings, checking for consistency between the problem descriptions and the canonical solutions, and ensuring that both the solutions and test cases were correct and efficient. Where inefficiencies were identified, suggestions for optimization were provided. The review also scrutinized the test cases to ensure comprehensive coverage, identifying any gaps that could allow incorrect solutions to pass. \minorrev{For the sake of completeness, we must note that the peer review process is exclusively concerned with evaluating the modifications made and does not encompass other elements of the procedure.}
While the initial creation of the benchmark took over 100 hours, the independent peer-review process required an additional 16 hours. As a result of this review, 1\% of the problems were redesigned due to structural issues, 9\% received additional test cases, and 15\% underwent minor grammar or clarity improvements. All suggested changes were documented and reviewed by the original author, with every recommendation either implemented or refined further upon discussion.
 Since the peer-review involved modifications beyond the test suites, completions for all models were re-run. Despite these changes, the results remained largely consistent with the original, with 40\% of models showing no change in their \texttt{pass@1} scores, 50\% showing a 1-2\% absolute change, and only 10\%—previously top performers—experiencing a 5\% drop. This demonstrated that the peer-review process upheld the benchmark's robustness while refining its quality.

\subsection{Experimental Setup}
\label{sec:humanevalpro:experiment}
\new{To assess the impact of the modifications applied to the benchmark, we examine the \textsc{pass@1} performance of ten state-of-the-art open-source software code models using the original HumanEval alongside two enhanced variants, \textsc{\humanevaladvanced} and \textsc{EvalPlus~\cite{HumanEvalPlus-Mini-MBPP+-2023}}.
We selected the top-performing models from the big code LLM leaderboards at the evaluation's start: ``NTQAI/Nxcode-CQ-7B-orpo'', ``Qwen/CodeQwen1.5-7B'', ``deepseek-ai/deepseek-coder-6.7b-instruct'', ``TechxGenus/starcoder2-15b-instruct'', ``ise-uiuc/Magicoder-S-DS-6.7B'', ``Artigenz/Artigenz-Coder-DS-6.7B'', ``HuggingFaceH4/starchat2-15b-v0.1'', ``google/codegemma-7b-it'', ``codeLlama/CodeLlama-13b-Instruct-hf'', and ``Stabilityai/stable-code-3b''.}
For each task in the benchmark (164 total), the LLM is prompted using an instructional preamble asking the model to finish the implementation of the function requested in addition to providing the imports, function header, and function description with each request.
 We run the inference for the models on a cluster with one NVIDIA A100 80GB GPU and 32 CPU cores. During test execution, a timeout limit of 15 seconds is utilized per function call to disregard completions that are potentially looping forever or are considered overly inefficient with regard to the canonical solutions. Each test is executed in an evaluation suite using the precautions deployed by OpenAI.

\subsection{Results}\label{sec:results}

\begin{figure}[tb]
    \centering
    \includegraphics[width=0.5\textwidth]{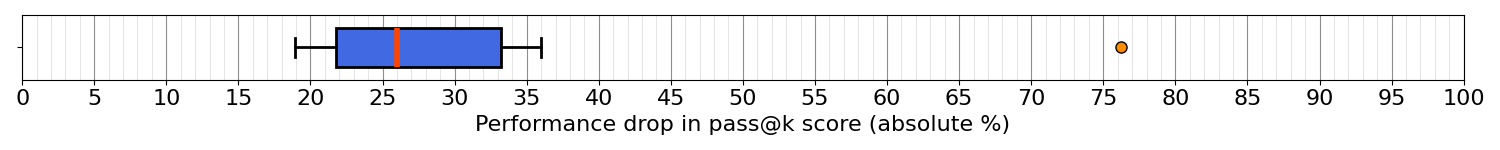}
    \caption{Boxplot depicting the distribution of absolute drops in \texttt{pass@1} score between HumanEval and the newly introduced \textsc{\humanevaladvanced} benchmark, based on 10 LLMs (\autoref{tab:humaneval_performance}).}
    \label{fig:humanevalpro:performance-drops-distribution}
\end{figure}

\new{Upon conducting the experiments outlined in \autoref{sec:humanevalpro:experiment}, a key observation is an \textbf{average decrease of 31.22\%} and a \textbf{median decrease of 26.02\%} (both absolute percentages) in pass@$1$ results of HumanEval when contrasted with the newly introduced \textsc{\humanevaladvanced} benchmark, using 10 different LLMs. Specific model outcomes are detailed in \autoref{tab:humaneval_performance}, with an overall depiction of the performance decreases displayed in \autoref{fig:humanevalpro:performance-drops-distribution}. While there are still significant performance declines when comparing HumanEvalPlus with the original HumanEval (11.28 mean and 7.05 median), the declines are substantially larger with \humanevaladvanced than with EvalPlus. Overall, the findings highlight a marked reduction in model performance when measured against \textsc{\humanevaladvanced} as opposed to the initial HumanEval benchmark. This pattern emphasizes the enhanced difficulty and refined evaluation precision offered by \textsc{\humanevaladvanced}, which incorporates more resilient environments featuring type annotations and clearer instructions.}

 A closer look reveals that the top-performing models from the original HumanEval benchmark do not maintain their standings in \textsc{\humanevaladvanced}. For example, while \textsc{\humanevaladvanced} consistently ranks \texttt{deepseek-ai}'s \texttt{deepseek-coder-6.7b-instruct} as the top performer, previous leaders like \texttt{NTQAI}'s \texttt{Nxcode-CQ-7B-orpo} and \texttt{Qwen}'s \texttt{CodeQwen1.5-7B} show a significant drop in their rankings. Specifically, \texttt{NTQAI}'s \texttt{Nxcode-CQ-7B-orpo} falls from an impressive 87.23\% \texttt{pass@1} in HumanEval to 51.22\% in \textsc{\humanevaladvanced}, and \texttt{Qwen}'s \texttt{CodeQwen1.5-7B} plummets from 87.2\% to 10.98\%. This sharp decline suggests that certain models may have benefited from data leakage or other issues in the original HumanEval benchmark, indicating the necessity for a more rigorous benchmark like \textsc{\humanevaladvanced} to accurately assess model performance.

 Especially the resilience of models, e.g., \texttt{deepseek-ai/deepseek-coder-6.7b-instruct}, can be confirmed by evaluating the model on the new challenges presented in \textsc{\humanevaladvanced}. When models also score relatively well in this benchmark, it highlights the models' ability to adapt and perform competently under more demanding conditions, making \textsc{\humanevaladvanced} a great benchmark to reveal the reliability of the capabilities of models. This finding also emphasizes the need for regular updates to benchmarks, as reliance on outdated benchmarks can potentially misrepresent model capabilities over time, even when models claim not to train on the test data.

 Furthermore, an interesting observation emerges when analyzing model size and performance: bigger is not always better. Larger models such as \texttt{TechxGenus/starcoder2-15b-instruct} and \texttt{HuggingFaceH4/starchat2-15b-v0.1} do not consistently outperform smaller models. This observation suggests that model size alone is not a definitive predictor of success in complex, edge-case-inclusive benchmarks like \textsc{\humanevaladvanced}. For instance, looking at \texttt{pass@1} scores, \texttt{Techx-} \texttt{Genus/starcoder2-15b-instruct} scores 77.4\% on HumanEval but drops to 43.29\% on \textsc{\humanevaladvanced}, while \texttt{ise-uiuc/Magicoder-S-DS-6.7B} scores 76.8\% on HumanEval (lower) but only drops to 53.66\% on \textsc{\humanevaladvanced} (higher). This highlights that increased model size does not necessarily equate to better performance in more challenging assessments.

Lastly, ranking problems based on their difficulty using pass metrics per problem, confirms that \textsc{\humanevaladvanced} presents a well-distributed range of complexity over the complete set of challenges. It also shows the increased difficulty of the benchmark, where even the easiest problems are not universally passed, with roughly 30\% of the models still failing to solve them. Altogether, this distribution highlights \textsc{\humanevaladvanced}'s effectiveness in providing a thorough assessment environment, making it an excellent tool for evaluating straightforward coding capabilities of LLMs in a lightweight manner, the main reason behind the popularity of the original HumanEval benchmark and its variants.

\begin{table}[h!]
\centering
\caption{Performance Comparison on HumanEval Benchmarks (pass@1) - Values in parentheses are the $\Delta$ with the baseline HumanEval.}
\label{tab:humaneval_performance}
\begin{tabular}{lccc}
\toprule
\textbf{Model} & \textbf{HumanEval} & \textbf{HumanEval+} & \textbf{HumanEvalNext} \\
\midrule
stable-code & 30.72 & 25.60 ($-5.12$) & 1.83 ($-28.89$) \\
CodeLlama & 50.60 & 34.10 ($-16.50$) & 29.88 ($-20.72$) \\
codegemma & 60.40 & 51.80 ($-8.60$) & 41.46 ($-18.94$) \\
starchat2 & 73.80 & 71.30 ($-2.50$) & 43.29 ($-30.51$) \\
Artigenz-Coder & 75.60 & 72.60 ($-3.00$) & 53.66 ($-21.94$) \\
Magicoder & 76.80 & 71.30 ($-5.50$) & 53.66 ($-23.14$) \\
starcoder2 & 77.40 & 60.00 ($-17.40$) & 43.29 ($-34.11$) \\
deepseek-coder & 80.22 & 71.30 ($-8.92$) & 58.54 ($-21.68$) \\
CodeQwen1.5 & 87.20 & 45.70 ($-41.50$) & 10.98 ($-76.22$) \\
Nxcode-CQ & 87.23 & 83.50 ($-3.73$) & 51.22 ($-36.01$) \\
\bottomrule
\end{tabular}
\end{table}

\begin{table}[h!]
\centering
\caption{Performance Comparison on HumanEval Benchmarks (pass@1) - Values in parentheses are the $\Delta$ with the baseline HumanEval.}
\label{tab:humaneval_performance}
\begin{tabular}{lccc}
\toprule
\textbf{Model} & \textbf{HumanEval} & \textbf{HumanEval+} & \textbf{HumanEvalNext} \\
\midrule
stable-code & 30.72 & 25.60 ($-5.12$) & 1.83 ($-28.89$) \\
CodeLlama & 50.60 & 34.10 ($-16.50$) & 29.88 ($-20.72$) \\
codegemma & 60.40 & 51.80 ($-8.60$) & 41.46 ($-18.94$) \\
starchat2 & 73.80 & 71.30 ($-2.50$) & 43.29 ($-30.51$) \\
Artigenz-Coder & 75.60 & 72.60 ($-3.00$) & 53.66 ($-21.94$) \\
Magicoder & 76.80 & 71.30 ($-5.50$) & 53.66 ($-23.14$) \\
starcoder2 & 77.40 & 60.00 ($-17.40$) & 43.29 ($-34.11$) \\
deepseek-instruct & 80.22 & 71.30 ($-8.92$) & 58.54 ($-21.68$) \\
CodeQwen1.5 & 87.20 & 45.70 ($-41.50$) & 10.98 ($-76.22$) \\
Nxcode-CQ & 87.23 & 83.50 ($-3.73$) & 51.22 ($-36.01$) \\
\bottomrule
\end{tabular}
\end{table}

\section{Discussion}
\label{sec:discussion}
\rohamrev{
In this section, we discuss the broader implications of our study; here, we attempt to tie our findins back to the RQs that guided our study. We, specifically, touch on what our results mean for researchers navigating the field, address the practical salability of our proposed approach, and outline threats to validity and opportunities for future work.
}

\subsection{Implications}
\label{sec:implications}
\subsubsection{Finding the Right Benchmark}
\rohamrev{
In our analysis, and specifically RQ1 we find that AI4SE benchmarking is highly fragmented, with many benchmarks suffering from poor maintenance and lack of discoverability. With this finding, we underscore the fact a major challenge in AI4SE is the difficulty of selecting a contextually relvant benchmark. \searchtool was our direct answer to RQ2. 
}
With \searchtool, users can select more relevant benchmarks and gain deeper insights into model performance for their specific needs.
\subsubsection{Benchmarking the Benchmark}
\rohamrev{
Our answer to RQ3, \benchframework, can have an impact on how we measure progress in AI4SE.} Our results show the significant impact of integrating the \textsc{\benchframework} approach and highlight the value of peer-reviewed, validated benchmarks. With many state-of-the-art models tested on similar benchmarks, both industry and academia should adapt their evaluation methods to ensure robust results.
The refinement of benchmarks across various AI4SE tasks will be critical for guiding future research and ensuring that these models can perform effectively.
\new{Although this study, \rohamrev{specifically in RQ1,} highlights the issue of data leakage in current benchmarks, it remains true that \humanevaladvanced would also be affected by this issue. However, we argue that benchmarks should not be used indefinitely and should evolve to pose increasing challenges in alignment with model improvements.}

\subsection{A Case Study on the agentification of \benchframework}
\label{sec:case-study-agent}
A potential criticism of the approach proposed in RQ3 is the significant manual effort it demands compared to a more automated system.
To address these concerns about the generalizability and manual labor intensity of BenchFrame, we developed an agentic pipeline designed to replicate and automate the benchmark improvement process with minimal human intervention. Our baseline for this effort was the HumanEvalNext dataset, which had previously been improved by manual human supervision. We aimed to determine whether an agentic approach could yield comparable improvements without the extensive human labor (to show noninfiriority).

The agent pipeline was structured into three distinct phases, which mirrors the manual process described in \autoref{sec:humanevalpro:approach:modifications}. In the first phase, the agent improved the problem description, which includes the function signature and the docstring. Using this refined description, the agent then generated an improved canonical solution. In the final phase, the agent created and set up the test cases to validate the solution. This process was applied across all the problems in the dataset, and our hypothesis was that this automated approach would be noninferior to the human-improved dataset.

To validate this, we conducted a paired evaluation comparing the human- and agent-improved versions of each benchmark problem in the HumanEvalNext dataset, specifically \humanevaladvanced versus \humanevaladvancedagentic. Two independent reviewers rated each pair using a five-point ordinal scale ranging from -2 (strongly preferring the human version) to +2 (strongly preferring the agentic version), with 0 indicating no preference. We tested whether the agentic pipeline was noninferior to the human-improved process, using a conservative noninferiority margin of $\delta$ = -0.5. That is, we assessed whether the average rating for the agentic outputs was not more than 0.5 points worse than the human outputs.

For each reviewer, as well as for the combined ratings (averaged across reviewers), we performed both a one-sided one-sample t-test and a Wilcoxon signed-rank test to assess noninferiority. The results were consistent and conclusive: the mean ratings were 0.16 (SD = 0.67) for Reviewer 1, 0.53 (SD = 0.51) for Reviewer 2, and 0.35 (SD = 0.49) for the combined average. The corresponding t-statistics were 12.68, 25.75, and 22.10, with one-sided p-values all equal to 1.000 (indicating extremely strong evidence in favor of noninferiority given the direction of the test). The non-parametric Wilcoxon signed-rank tests further confirmed these results, with statistics of 12,155.5 (p = $3.46\times10^{-20}$), 13,198.0 (p = $7.20\times10^{-28}$), and 12,698.0 (p = $5.94\times10^{-27}$) respectively.

These results indicate that the agentic pipeline is noninferior to the human-improved process—indeed, with mean and median ratings above zero, the agentic output was often preferred. This supports the feasibility of using an agentic approach to automate benchmark refinement while significantly reducing manual effort.

While noninferior, we noted several standardized observations, in terms of pitfalls, during this evaluation, such as (1) instances where the agent's docstring occasionally revealed aspects of the solution, or (2) where canonical solutions made unintended assumptions, or (3) where the test cases did not adhere to the required, single-line assert, format. These observations were later accounted for in later stages of the feasibility study to make adjustments in the pipeline, and to ensure that it remained flexible and effective.

\begin{figure}[tb]
    \centering
    \resizebox{\columnwidth}{!}{
    \begin{tikzpicture}[node distance=0.3cm]
        \rohamrev{
        \tikzset{
          arrow/.style={-stealth,thick},
          dashedarrow/.style={arrow,dashed}
        }

        \node (start) [startstop] {};
        \node (init) [process, right=of start] {Initialize attempt = 1};
        \node (textimp) [process, right=of init] {Text Improvement (a)};
        \node (codeimp) [process, right=of textimp] {Code Improvement (b)};
        \node (testimp) [process, right=of codeimp] {Test Improvement (c)};
        \node (check) [decision, below=of codeimp, yshift=-0.25cm] {All tests pass?};
        \node (validate) [process, right=of check, xshift=0.72cm] {Validation (d)};
        \node (done) [startstop, above=of check] {};

        \node (attemptlt3) [decision, below=of textimp, yshift=-0.3cm] {attempt $\leq$ 3};
        \node (incr) [process, below=of init, yshift=-0.4cm] {attempt = attempt + 1};
        \node (findpass) [process, below=of check, yshift=0.16cm, xshift=2.5cm] {Minimized passing test cases (e)};
        \node (fail) [process, below=of attemptlt3] {Mark task as failed};

        \draw[arrow] (start) -- (init);
        \draw[arrow] (init) -- (textimp);
        \draw[arrow] (textimp) -- (codeimp);
        \draw[arrow] (codeimp) -- (testimp);
        \draw[arrow] (testimp) -- (validate);
        \draw[arrow] (validate) -- (check);
        \draw[arrow] (check.north) -- node[left, font=\scriptsize] {yes} (done);
        \draw[arrow] (check.west) -- node[above, font=\scriptsize] {no} (attemptlt3);
        \draw[arrow] (attemptlt3.west) -- node[above, font=\scriptsize] {yes} (incr);
        \draw[arrow] (attemptlt3.south east) -- node[above, font=\scriptsize] {no} (findpass.west);
        \draw[arrow] (incr.north) to[out=90,in=225] (textimp.south west);

        \draw[dashedarrow] (findpass.north) to[out=120,in=-20] (check.east);
        \draw[dashedarrow] (check.north east) to[out=40,in=0] node[right, font=\scriptsize] {yes} (done);
        \draw[dashedarrow] (check.south) to[out=-90,in=0] node[below, font=\scriptsize] {no} (fail.east);
        }
    \end{tikzpicture}
    }
    \caption{\rohamrev{The flow-chart indicating the pipeline used to improve MBPP.}}
    \label{fig:pipeline-agentic}
\end{figure}

For the sake of completeness, we present the refined pipeline used in the process of benchmark refinement. The pipeline begins by initializing an attempt counter, which tracks the number of iterations taken to complete a given benchmark improvement. The process then proceeds through three core phases: (a) Text Improvement, in which the agent refines the problem description and adds a type-annotated function signature; (b) Code Improvement, where the canonical implementation is modified according to Python best practices and based on the improved task description; and (c) Test Improvement, which involves generating assert-based test cases that include edge cases, boundary conditions, and common error scenarios. These three components are then assembled into a complete Python program and passed to the Validation phase (d), which runs the implementation against the test suite in a sandboxed subprocess.

If all tests pass on the first try, the process terminates successfully. Otherwise, the pipeline evaluates whether the number of attempts is below a fixed threshold (three, in our case). If so, the pipeline either increments the attempt counter and retries the full process or, alternatively, in the case that more tests are passing than were present in the original test suite, the pipeline identifies a minimal set of passing tests (e) to be used as the new test suite. This loop ensures that even partially successful test runs can be used to improve the benchmark iteratively. After three failed attempts, the task is marked as unsuccessful.

For our experiments, we used the OpenAI API \footnote{\url{https://platform.openai.com}}, specifically, the \texttt{o3-mini-2025-01-31} as the underlying model for the agent. We evaluated this new dataset on the same set of models, and the results (as presented in \autoref{tab:humanevalnextagentic:results}) indicated comparable (and even better) performance patterns. These findings provide adequate justification for how an agentic pipeline can effectively generalize and provide a semi-automated solution for benchmark improvement.

\begin{table}[tb]
\centering
\caption{\rohamrev{Pass@1 results for HumanEval, \humanevaladvanced, and \textsc{\humanevaladvancedagentic}.}}
\label{tab:humanevalnextagentic:results}
\begin{tabular}{p{18mm}cp{19mm}p{19mm}}
\toprule
\textbf{Model} & \textbf{HumanEval} & \textbf{\humanevaladvancedagentic ($\Delta$)} & \textbf{\humanevaladvanced ($\Delta$)} \\
\midrule
stable-code & 30.72 & 2.44($-28.28$) & 1.83 ($-28.89$) \\
CodeLlama & 50.60 & 18.90($-31.70$) & 29.88 ($-20.72$) \\
codegemma & 60.40 & 22.56 ($-37.84$) & 41.46 ($-18.94$) \\
starchat2 & 73.80 & 26.83 ($-46.97$) & 43.29 ($-30.51$) \\
Artigenz-Coder & 75.60 & 31.70 ($-43.90$) & 53.66 ($-21.94$) \\
Magicoder & 76.80 & 31.70 ($-45.10$) & 53.66 ($-23.14$) \\
starcoder2& 77.40 & 14.02 ($-63.38$) & 43.29 ($-34.11$) \\
deepseek-coder& 80.22 & 35.37 ($-44.85$) & 58.54 ($-21.68$) \\
CodeQwen1.5 & 87.20 & 1.22 ($-86.00$) & 10.98 ($-76.22$) \\
Nxcode-CQ & 87.23 & 31.70 ($-55.53$) & 51.22 ($-36.01$) \\
\bottomrule
\end{tabular}
\end{table}

The results presented indicate, that while a certain level of human intervention, specifically in the review phase, is still required to ensure quality of the improvements, it is but a fraction of the total time required to manually improve the benchmark. From a cost perspective, we can also report that the total incurred cost from calling the apis for the models was \$5.28\footnote{At the time of running these experiments, on the 31st of July 2025}. This is but a small portion of the labor costs required to improve the benchmark manually. The introduction of this agentic layer significantly reduced the manual effort required. We claim that, with this approach, one can extend it to other datasets with similar success.

\subsection{Evaluating the Generalizability of \benchframework}
An additional critique that can be raised regarding our approach is its generalizability. To tackle this issue, we adopted a strategy akin to HumanEval. Using our systematic observations of MBPP from the review, we apply \benchframework to 100 problems selected from a pool of 500 MBPP test samples. The rationale behind this selection is twofold: firstly, it aims to assess a new task, specifically program synthesis, and secondly, it considers the benchmark's popularity. During this process, we subject the function descriptions, implementations (including variable typing), and significantly extend test cases; more specifically, we go from a fixed 3 tests per problem to an average of 12.35, with a median of 11.5. We finally go through several rounds of peer review to ensure the quality of the improvements. We refer to the resulting benchmark as MBPPNext. The evaluation results of this dataset in comparison to the baseline are detailed in~\autoref{tab:mbppnext:results}. We observe a comparable decline in performance similar to that noted for HumanEval, specifically, an average decrease of 13.4 percentage points across the methods evaluated. These findings further underscore the necessity for thorough peer review and verification of benchmarks, as discussed in previous sections of this paper.

\begin{table}
\centering
\caption{Pass@1 Results for MBPP and MBPPNext. Based on \url{https://github.com/google-research/google-research/blob/master/mbpp/mbpp.jsonl}~{\texttt{accessed on 14-07-2025}}}
\label{tab:mbppnext:results}
\begin{tabular}{lcc}
\toprule
\textbf{Model} & \textbf{MBPP} & \textbf{MBPPNext ($\Delta$)} \\
\midrule
stable-code & 34.0 & 28.0 (-6\%) \\
CodeLlama & 40.0 & 31.0 (-9\%) \\
codegemma & 43.0 & 33.0 (-10\%) \\
starchat2 & 59.0 & 42.0 (-17\%) \\
Artigenz-Coder & 60.0 & 47.0 (-13\%) \\
Magicoder & 63.0 & 45.0 (-18\%) \\
starcoder2 & 44.0 & 36.0 (-8\%) \\
deepseek-coder & 68.0 & 50.0 (-18\%) \\
CodeQwen1.5 & 53.0 & 40.0 (-13\%) \\
Nxcode-CQ & 74.0 & 52.0 (-22\%) \\
\bottomrule
\end{tabular}
\end{table}

\subsection{Future Work}
\label{sec:future_work}
Future work on \textsc{\benchframework} will focus on expanding to additional programming languages, which it already has been optimized for, yet these variants have not been produced or evaluated. 
Although evaluating \modelcount LLMs provides valuable insight, it remains unclear how larger, top-performing models behind paywalls, such as GPT and Gemini, would perform; Future research could evaluate the performance differences of such models when comparing the base and the pro version. \rohamrev{Furthermore, future research should focus on applying the underlying ideas of \benchframework to mutli-file and project-level benchmarks like Defects4J.}

\subsection{Threats to the Validity}\label{sec:threats}

\textbf{Construct Validity}:
To reduce bias and errors in the literature review, two authors followed a structured protocol for selecting and filtering sources. A threat to construct validity comes from the subjectivity in defining and applying the inclusion/exclusion criteria. To address this, we pre-defined clear criteria and peer-reviewed the selection process for consistency.
Another potential validity threat stems from the design of our user study, as it may not fully capture the tool's overall functionality, strengths, or weaknesses. To mitigate this, we evaluated diverse aspects including usability, functionality, usefulness, and intuitiveness.
Lastly, to minimize biases during \humanevaladvanced's development, we applied consistent refinement criteria and peer-reviews, although subjective interpretation remains a residual risk.

\textbf{Internal Validity}: 
A potential threat in the user study is selection bias. To mitigate this, we included participants from both industry and academia, ensuring a range of skills and experience. \rohamrev{To reduce the real-world impact arising from any study biases, we have created a continuous feedback mechanism in \searchtool to receive continuous feedback from users to be able to meet their needs.}
All participants received the same tool, guidance, and instructions. To reduce bias in the peer review or \humanevaladvanced, the reviewer was not informed of specific changes made by the first author. 

\textbf{External Validity}:
A threat here is the generalizability of the user study's results. We mitigated this by including \usercount participants, but the sample size may still limit the generalizability of the results.
Although we evaluated the effects of the \textsc{\benchframework} approach on the performance of \modelcount models with one of the most widely used benchmarks, this could not be considered sufficient to prove the generalizability of the results; adding more models and application to more benchmarks could further confirm our results.
\new{\benchframework may be criticized for its limited applicability due to the significant manual effort it demands compared to more automated systems like EvalPlus. \rohamrev{We demonstrate that the process of improving benchmarks such as HumanEval and MBPP can be (semi-)automated through the use of agentic pipelines.} 
}

\rohamrev{
\section{Related Work}
\label{sec:rel-work}
The rapid integration of LLMs into software engineering has led to a corresponding proliferation of benchmarks designed to evaluate them. This has spurred research to systematically map this new terrain, assess the quality of evaluation resources, and build tools for navigating them. Our work is situated within these three emerging areas.

\paragraph{AI4SE Surveys and Taxonomies}
Initial research in this area included broad surveys of LLMs in software engineering, which established evaluation as a central research challenge~\cite{fan2023largelanguagemodelssoftware,hou2024largelanguagemodelssoftware,zakerinasrabadi2023systematicliteraturereviewsource,quanjun_zhang_critical_2023}. More recently, meta-analyses have focused on the benchmarks themselves. For instance, Wang et al. created a taxonomy based on the Software Development Life Cycle, revealing a research gap with a strong focus on code implementation and a negligence towards design and requirements engineering~\cite{wang2025softwaredevelopmentlifecycle}. Our work contributes to this area by providing a comprehensive review of \benchcount benchmarks from \papercount studies (\textbf{RQ1}). We go beyond categorization by analyzing each benchmark across a 14-point metadata schema to identify systemic limitations, such as poor maintenance, language specificity, and a lack of peer review.

\paragraph{Tools for Benchmark Discovery and Navigation}

While platforms like Hugging Face and the now-discontinued Papers With Code serve as valuable repositories for hosting datasets~\cite{huggingface}, they often lack the specialized search and visualization capabilities needed by AI4SE researchers. \searchtool addresses this gap (\textbf{RQ2}) by providing an extensible semantic search tool specifically designed for the AI4SE community. Unlike general-purpose repositories, it combines semantic embeddings with structured metadata filters and an interactive visualization of the benchmark landscape to improve the discoverability and selection of relevant evaluation tools.

\paragraph{Methodologies for Improving Benchmark Quality}

There is a growing agreement that many static benchmarks are prone to saturation and data contamination, which can compromise the validity of evaluation results~\cite{jain_livecodebench_2024}. In response, two movements have arisen. Efforts like LiveCodeBench fall under the "build anew" category; these strive to develop dynamic, "live" benchmarks that continuously gather new challenges from real-world environments to avoid overfitting.

Our method, \benchframework, belongs to the "repair and refine" category and focuses on improving the quality of existing benchmarks (\textbf{RQ3}). A method within this category is DyCodeEval, which uses seed contexts to frame problems in specific scenarios (e.g., banking, healthcare, education). This approach is designed to produce more reliable results by minimizing data contamination and memorization effects~\cite{chen2025dynamicbenchmarkingreasoningcapabilities}. Our approach sets itself apart by incorporating a peer-review-oriented process that includes standardized observations, precise modifications, and independent validation. Like DyCodeEval, we show its practical scalability through an agentic pipeline that automates the improvement process and tackles the main challenge, namely, the manual efforts required in \benchframework.
}

\section{Conclusion}
\label{sec:conclusion}
The findings of our study highlight the importance of reliable and consistent benchmarking in AI4SE to drive the development of more robust models. Through the creation of \benchframework and the enhancement of the HumanEval benchmark, we have demonstrated that higher-quality benchmarks reveal substantial performance gaps, as shown by the 31.2\% average reduction in pass@1 scores across ten state-of-the-art models. This significant decline highlights the impact of more stringent evaluations. \searchtool further enhances this process by facilitating the discovery of relevant benchmarks and reduces the overhead associated with selecting the appropriate tools for evaluation.

\subsection{Data Availability}
We publicly release the results of our literature review, user study, and 50\% of the manually refined benchmark.~\footnote{\url{https://github.com/AISE-TUDelft/AI4SE-benchmarks}}
Upon acceptance, the complete benchmark will be made available on both GitHub and HuggingFace.

\section{Acknowledgments}
This research was supported in part by an Amazon Research Award granted to Dr. Maliheh Izadi. We gratefully acknowledge Amazon's support. The views and conclusions contained in this paper are those of the authors and do not necessarily reflect the position or policies of Amazon.

\bibliographystyle{IEEEtran}
\bibliography{main}

\end{document}